\documentclass[prapplied, reprint, superscriptaddress, floatfix, longbibliography]{revtex4-2}

\usepackage{mathtools}
\usepackage{textgreek}
\usepackage{graphicx}
\usepackage{bm}
\usepackage{hyperref}
\usepackage{braket}
\usepackage{xcolor}
\usepackage{multirow}
\usepackage{placeins}
\usepackage{mleftright}
\usepackage{pifont}
\usepackage{enumitem}
\usepackage[norefs, nocites, ignoreunlbld]{refcheck}

\begin{document}

\title{Generation of frequency-bin-encoded dual-rail cluster states via time-frequency multiplexing of microwave photonic qubits}

\author{Zhiling~Wang}
\email{zhiling.wang@riken.jp}
\affiliation{RIKEN Center for Quantum Computing (RQC), Wako, Saitama 351-0198, Japan}
\author{Takeaki~Miyamura}
\affiliation{Department of Applied Physics, Graduate School of Engineering, The University of Tokyo, Bunkyo-ku, Tokyo 113-8656, Japan}
\author{Yoshiki Sunada}
\affiliation{Department of Applied Physics, Stanford University, Stanford, California 94305, USA}
\author{Keika~Sunada}
\affiliation{Department of Applied Physics, Graduate School of Engineering, The University of Tokyo, Bunkyo-ku, Tokyo 113-8656, Japan}
\author{Jesper~Ilves}
\affiliation{Department of Applied Physics, Graduate School of Engineering, The University of Tokyo, Bunkyo-ku, Tokyo 113-8656, Japan}
\author{Kohei~Matsuura}
\affiliation{Department of Applied Physics, Graduate School of Engineering, The University of Tokyo, Bunkyo-ku, Tokyo 113-8656, Japan}
\author{Yasunobu~Nakamura}
\affiliation{RIKEN Center for Quantum Computing (RQC), Wako, Saitama 351-0198, Japan}
\affiliation{Department of Applied Physics, Graduate School of Engineering, The University of Tokyo, Bunkyo-ku, Tokyo 113-8656, Japan}

\date{\today}

\begin{abstract}
Cluster states are a class of multi-qubit entangled states with broad applications such as quantum metrology and one-way quantum computing. Here, we present a protocol to generate frequency-bin-encoded dual-rail cluster states using a superconducting circuit consisting of a fixed-frequency transmon qubit, a resonator and a Purcell filter. We implement time-frequency multiplexing by sequentially emitting co-propagating microwave photons, where each time bin contains a dual-rail qubit encoded in two distinct frequency modes, and adjacent time bins are entangled to form the cluster state. The frequency-bin dual-rail encoding enables erasure detection based on photon occupancy. We characterize the state fidelity using quantum tomography and quantify the multipartite entanglement using localizable entanglement. Our implementation achieves a state fidelity exceeding 50$\%$ for a cluster state consisting of up to four logical qubits. The localizable entanglement remains across chains of up to seven logical qubits. After discarding the erasure errors, the fidelity exceeds 50$\%$ for states with up to eight logical qubits, and the entanglement persists across chains of up to eleven qubits. These results highlight the improved robustness of frequency-bin dual-rail encoding against photon loss compared to conventional single-rail schemes. This work provides a scalable pathway toward high-dimensional entangled state generation and photonic quantum information processing in the microwave domain.
\end{abstract}

\maketitle

\section{INTRODUCTION\label{sec:intro}}
Quantum entanglement is a fundamental concept of quantum physics and an essential resource for quantum computing and quantum information processing~\cite{guhne2009entanglement,kempe1999multiparticle,horodecki2009quantum}. Among various entangled states, cluster states~\cite{briegel2001persistent}, a special class of graph states, serve as a versatile platform for measurement-based quantum computation~\cite{raussendorf2001one,raussendorf2003measurement,briegel2009measurement}, fault-tolerant error correction~\cite{schlingemann2001quantum,bell2014experimentalerrcor}, quantum metrology~\cite{friis2017flexible,shettell2020graph}, quantum secret sharing~\cite{markham2008graph,bell2014experimentalsharing}, and quantum repeaters~\cite{azuma2015all}. Recent advances have demonstrated propagating cluster states in both microwave and optical domains~\cite{kuznetsova2012cluster,bell2014experimentalerrcor,bell2014experimentalsharing,schwartz2016deterministic,besse2020realizing,istrati2020sequential,thomas2022efficient,cogan2023deterministic,banic2025exact}. However, these implementations typically rely on single-rail encoding, where the presence or absence of a photon defines a qubit. This makes them inherently vulnerable to photon loss during propagation.

To overcome this limitation, researchers have explored various alternative encoding schemes, such as time-bin encoding~\cite{kurpiers2019quantum,ilves2020demand}, temporal mode encoding~\cite{penas2024multiplexed}, path encoding~\cite{kannan2020generating}, and frequency-bin encoding~\cite{yang2024deterministic}. Among these, frequency-bin encoding offers a unique advantage: the ability to utilize multiple modes within a single temporal window, enabling compact and scalable entanglement generation. In particular, frequency-bin encoding naturally supports a dual-rail configuration, where each logical qubit is defined by the presence of a single photon across a pair of frequency channels. This dual-rail encoding enables photon-loss detection during propagation and aligns with recent advances in erasure detection in superconducting systems, which have demonstrated the utility of such encoding for error-resilient quantum information processing~\cite{grassl1997codes,knill2005scalable,teoh2023dual,mehta2025bias,huang2025logical,chou2024superconducting,levine2024demonstrating,koottandavida2024erasure}.

In this work, we demonstrate the generation of frequency-bin-encoded microwave photonic cluster states using a superconducting circuit composed of a fixed-frequency transmon qubit, a resonator and a Purcell filter. By sequentially emitting pairs of co-propagating microwave photons in two fixed, distinct frequency modes across multiple time bins, we realize linear cluster states comprising up to four logical qubits using the frequency-bin dual-rail encoding. We further infer the generation of larger logical cluster states based on the experimentally obtained process matrix. We perform quantum state tomography and evaluate localizable entanglement (LE) to characterize the state fidelity and robustness of the generated states. Without any error detection, we achieve a four-logical-qubit cluster state with fidelity exceeding 50$\%$ and LE extending across chains of up to seven logical qubits. When photon-loss detection is considered by focusing on the subspace spanned by the dual-rail basis states without loss, a similar fidelity can be achieved for an eight-logical-qubit cluster state, and LE can be observed over chains of up to eleven logical qubits.

\section{Frequency-Bin Encoding and Dual-Rail Cluster-State Generation\label{sec:method}}
\begin{figure*}[t]
    \centering
\includegraphics{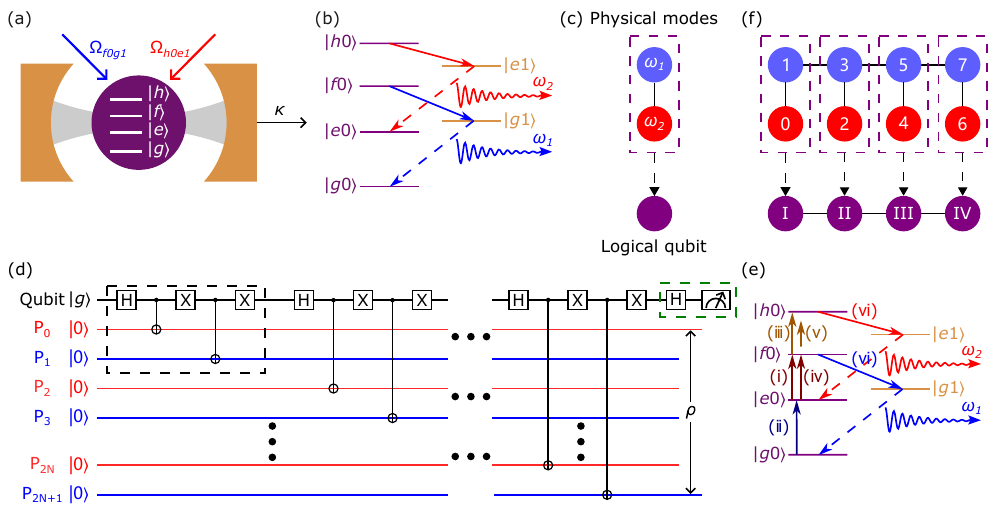}
    \caption{Protocol for generating frequency-bin photons and dual-rail cluster states. (a)~Schematic diagram of the system. The resonator--filter system has been simplified as an effective resonator. $\Omega_{f0g1}$ and $\Omega_{h0e1}$ represent the external drives for the $\ket{f0}$--$\ket{g1}$ and $\ket{h0}$--$\ket{e1}$ transitions, respectively. (b)~Energy-level diagram of the system. The two transitions we drive, $\ket{f0}$--$\ket{g1}$ and $\ket{h0}$--$\ket{e1}$ transitions, are shown as solid arrows. (c)~Frequency-bin dual-rail encoding. A logical qubit is defined by regarding the exclusive single-photon occupation of the frequency mode $\ket{\omega_1}$ ($\ket{\omega_2}$) as logical $\ket{0}_\mathrm{L}$~($\ket{1}_\mathrm{L}$). (d)~Quantum circuit used to generate a dual-rail frequency-bin cluster state. The part of the circuit enclosed by the black dashed box generates one frequency-bin photon pair. Here, H represents the Hadamard gate, and X represents the X gate~(the bit-flip gate). Finally, at the end of the sequence, the qubit is projected on the $X$-axis by applying a Hadamard gate and $Z$-axis measurement (enclosed by a green dashed box). (e)~Pulse sequence which is equivalent to the black dashed box in~(c): (i)~a $\pi_{ef}$ pulse, (ii)~a $\pi_{ge}$ pulse, (iii)~a $\pi_{fh}$ pulse, (iv)~a $\pi_{ef}$ pulse, (v)~a $\pi_{fh}/2$ pulse, and (vi)~two simultaneous pulses for driving the $\ket{f0}$--$\ket{g1}$ and $\ket{h0}$--$\ket{e1}$ transitions. (f)~Graph representation of the generated dual-rail frequency-bin cluster state. Here, as an example, we show a state with four dual-rail logical qubits. The colors of the vertices represent different frequency channels. The number on each mode corresponds to its order in the bra--ket representation. It can be regarded as a 1D cluster state (the mode order is notated using Roman numerals) under frequency-bin encoding.}
    \label{fig:scheme}
\end{figure*}

We realize this protocol using a circuit quantum electrodynamics (cQED) system consisting of a fixed-frequency transmon qubit and a resonator which are dispersively coupled with the transverse coupling strength $g$, as shown in Fig.~\ref{fig:scheme}(a). The resonator is also coupled through a Purcell filter to an output waveguide with a coupling rate~$\kappa$. Figure~\ref{fig:scheme}(b) shows the energy diagram of the system. Here, $\{\ket{g},\ket{e},\ket{f},\ket{h}\}$ represents the ground state and the first, second, and third excited states of the qubit, respectively, and $\{\ket{0},\ket{1}\}$ represents the Fock state of the resonator. 

Raman-type photon emission has been demonstrated in such circuit-QED systems using external drives~\cite{pechal2014,zeytinoglu2015}. These demonstrations typically use the transition between $\ket{f0}$ and $\ket{g1}$ states, which we refer to as the $\ket{f0}$--$\ket{g1}$ transition. In our protocol, we also exploit the analogous $\ket{h0}$--$\ket{e1}$ transition. As shown in Ref.~\citenum{miyamura2025}, the frequencies of the emitted photons by these two processes can be manipulated by tuning the frequencies of the external $\ket{f0}$--$\ket{g1}$ and $\ket{h0}$--$\ket{e1}$ drives, respectively. Thus, we can simultaneously generate two distinct photon modes with a desired frequency difference, as shown in Fig.~\ref{fig:scheme}(b).

By simultaneously applying two external drives for the $\ket{f0}$--$\ket{g1}$ and $\ket{h0}$--$\ket{e1}$ transitions, the superposition state between the qubit's $\ket{f}$ and $\ket{h}$ states can be mapped to a pair of co-propagating photon modes at different frequencies as follows:
\begin{equation}    \left(\alpha\ket{f}+\beta\ket{h}\right)\ket{00}\rightarrow\alpha\ket{g}\ket{01}+\beta\ket{e}\ket{10}.
\label{eq:two_freq_generate}
\end{equation}
Here $\ket{01}$ ($\ket{10}$) represents the state in which a photon at the frequency $\omega_1$ ($\omega_2$) exists in the emitted mode. Based on this co-propagating photon pair, we can realize a frequency-bin dual-rail encoding within the logical subspace spanned by $\ket{01}$ and $\ket{10}$, notating them as $\ket{\omega_1}$ and $\ket{\omega_{2}}$, respectively, as shown in Fig.~\ref{fig:scheme}(c).

As a whole system, the qubit state and the generated photons still maintain the entanglement between each other. Thus, this protocol also allows us to generate entangled states with multiple modes in the time domain by repeating the protocol~\cite{lindner2009proposal}. Figure~\ref{fig:scheme}(d) shows a quantum circuit to generate a dual-rail cluster state with time- and frequency-domain multiplexing. 
The core of this protocol is the sub-circuit enclosed by the black dashed box in Fig.~\ref{fig:scheme}(d), which is responsible for repeated photon-pair emission. An equivalent pulse sequence to realize this circuit is shown in Fig.~\ref{fig:scheme}(e). By using the four $\pi$ pulses marked with (i)--(iv) between the $\{\ket{g},\ket{e},\ket{f},\ket{h}\}$ levels, we map the qubit state from the $\{\ket{g},\ket{e}\}$ subspace into the $\{\ket{f},\ket{h}\}$ subspace. After that, we apply a $\pi_{fh}/2$ pulse~[marked as~(v)] and then simultaneously apply the two external drives~[marked as~(vi)]. The whole qubit--itinerant-photon system will be in the state in Eq.~\eqref{eq:two_freq_generate} and generate a single photon in a superposition state between two frequency modes~[P$_0$ and P$_1$ in Fig.~\ref{fig:scheme}(d)]. Details can be found in Appendix~\ref{app_sec:generation_seq}.

By repeating this generation sequence, we generate states with multiple logical qubits. For instance, after two rounds, the system state becomes (also shown in the ket notation used for the density matrix)
\begin{equation}
\begin{split}
    &\frac{1}{\sqrt{2}}\left[\ket{g}\ket{\omega_1}(\ket{\omega_1}+\ket{\omega_2})+\ket{e}\ket{\omega_2}(\ket{\omega_1}-\ket{\omega_2})\right]\\
    =&\frac{1}{\sqrt{2}}\left[\ket{g}\ket{01}(\ket{01}+\ket{10})+\ket{e}\ket{10}(\ket{01}-\ket{10})\right].
\end{split}    
\end{equation}
By projecting the qubit state to the $\ket{g}\pm\ket{e}$ basis~[the measurement in the circuit in Fig.~\ref{fig:scheme}(d)], we disentangle the qubit and photon modes and obtain an itinerant photonic state. By post-selecting the generated state based on the qubit measurement outcome in the $\ket{g}\pm\ket{e}$ basis, the corresponding photonic state becomes (also shown in the ket notation used for the density matrix)
\begin{equation}
\begin{split}
    \ket{\psi_\pm}&=\frac{1}{2}(\ket{\omega_1}\ket{\omega_1}+\ket{\omega_1}\ket{\omega_2}\pm\ket{\omega_2}\ket{\omega_1}\mp\ket{\omega_2}\ket{\omega_2})\\
    &=\frac{1}{2}(\ket{0101}+\ket{0110}\pm\ket{1001}\mp\ket{1010}).
    \label{eq:logical_two_cluster}
\end{split}
\end{equation}
Both generated states are locally equivalent to a 1D cluster state in the frequency-bin dual-rail encoding. In Fig.~\ref{fig:scheme}(f), we show the graph representation of a photonic state with four pairs of frequency modes, which is obtained by repeating the photon-emission protocol four times. Details of this equivalence can be found in Appendix~\ref{app_sec:graph_repre}. It can be regarded as a 1D cluster state based on the encoding scheme shown in Fig.~\ref{fig:scheme}(c).

\section{EXPERIMENT\label{sec:experiment}}
We implement the scheme introduced above using a device with the same structure as in Ref.~\citenum{miyamura2025}. The device consists of a fixed-frequency transmon qubit and two resonator--filter systems. In our experiment, only one of the resonator--filter systems is used. In the experiment, we use the first four energy levels $\{\ket{g},\ket{e},\ket{f},\ket{h}\}$ of the transmon, with transition frequencies $\omega_{ge}/2\pi=8021.8$~MHz, $\omega_{ef}/2\pi=7702.9$~MHz and $\omega_{fh}/2\pi=7347.6$~MHz. The qubit parameters were chosen to balance the Purcell limitation, Raman transition strength, and the coherence of higher-energy levels ~\cite{koch2007charge}.

For out-coupling, we achieve an effective resonator linewidth of $\kappa/2\pi=53.2$~MHz while maintaining a sufficiently long qubit energy-relaxation time by using a two-stage Purcell filter~\cite{sunada2022fast,miyamura2025}. Not only suppressing qubit decay through the resonator, this two-stage Purcell filter also enables the engineering of a broader effective coupling bandwidth, which allows a wider frequency separation between the emitted photon modes. The frequency of the resonator mode is $\omega_\mathrm{r}/2\pi=10300$~MHz. Other device parameters can be found in Appendix~\ref{app_sec:setup}.

\subsection{Spectra of generated photon states}

\begin{figure}
    \centering
\includegraphics{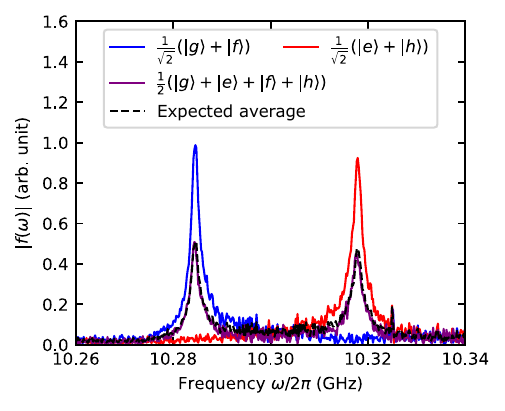}
\caption{Spectra of the emitted photon, $f(\omega)$, under different initial qubit states. The blue line corresponds to the qubit state prepared at $(\ket{g}+\ket{f})/\sqrt{2}$ state, and the red line corresponds to the qubit state prepared at $(\ket{e}+\ket{h})/\sqrt{2}$ state. They demonstrate that the frequency of the generated photons under two simultaneous drives depends only on the initially prepared qubit state. The purple line is the spectrum when the qubit is prepared in $(\ket{g}+\ket{e}+\ket{f}+\ket{h})/2$. The black dashed line is the expected average of the blue and red lines. All spectra are normalized by the same factor. Details can be found in Appendix~\ref{app_sub_sec:photon-generation-cali}.}
\label{fig:spectra}
\end{figure}

To confirm that we can simultaneously generate two co-propagating photon modes at desired frequencies, we measure the waveforms of the emitted photons from different qubit states. By preparing the initial qubit state at $(\ket{g}+\ket{f})/\sqrt{2}$ or $(\ket{e}+\ket{h})/\sqrt{2}$ and simultaneously driving $\ket{h0}$--$\ket{e1}$ and $\ket{f0}$--$\ket{g1}$ transitions, the emitted photon should be the $(\ket{0}+\ket{1})/\sqrt{2}$ state at the frequency of $\omega_{1}$ or $\omega_{2}$, respectively. Details can be found in Appendix~\ref{app_sub_sec:photon-generation-cali}. Figure~\ref{fig:spectra} shows the spectra of the emitted photons, $f(\omega)$, under different initial states. The spectra are obtained from the Fourier transformation of the measured waveform in the time domain. From the measurement results, we find that the center frequency of the photon mode $\ket{\omega_1}$ is $\omega_1/2\pi=10283$~MHz, and that of the photon mode $\ket{\omega_2}$ is $\omega_2/2\pi=10315$~MHz. 
The overlap between these two photon modes $\ket{\omega_1}$ and $\ket{\omega_2}$ represents the orthogonality between the two physical modes constituting a logical qubit, and can be calculated from their respective spectra, $f_{\omega_1}(\omega)$ and $f_{\omega_2}(\omega)$, as
\begin{equation}
    \braket{\omega_1,\omega_2}=\int f^*_{\omega_1}(\omega)f_{\omega_2}(\omega)\,d\omega.
\end{equation}
In our experiment, we choose an overlap of $3\%$ between the two photon modes with a duration of $1  \mathrm{\,\mu s}$ and a frequency difference of 32~MHz. This level of orthogonality is sufficient to ensure that the two modes are distinguishable for frequency-bin encoding. In principle, the two modes could be made fully orthogonal by further shaping the temporal envelopes and adjusting the relative drive phases to cancel their overlap. However, given that the dominant source of infidelity in our experiment originates from the qubit’s finite lifetime, the present 3$\%$ mode overlap has a negligible effect on the overall performance. The spectrum corresponding to the initial qubit state $(\ket{g}+\ket{e}+\ket{f}+\ket{h})/2$ is also shown in Fig.~\ref{fig:spectra}. The spectrum of this four-state superposition state shows the presence of both frequency components, while the process tomography results in Appendix~\ref{subsec:method_process_tomo} reveal the phase coherence between the two modes. Together, these observations indicate that we have generated a photon in the superposition of two modes at different frequencies.
Moreover, its spectral strength corresponds to the average (the black dashed line) of the blue [prepared at $(\ket{g}+\ket{f})/\sqrt{2}$] and red lines [prepared at $(\ket{e}+\ket{h})/\sqrt{2}$], indicating that the initial quantum state does not affect the two Raman processes. The photon generation efficiency is measured to be $96.9\%$ for the $\ket{f0}$--$\ket{g1}$ transition and $96.7\%$ for the $\ket{h0}$--$\ket{e1}$ transition. Details of the drive calibration can be found in Appendix~\ref{app_sec:cali}. 

\subsection{Generating dual-rail cluster states}

\begin{figure*}
    \centering
\includegraphics{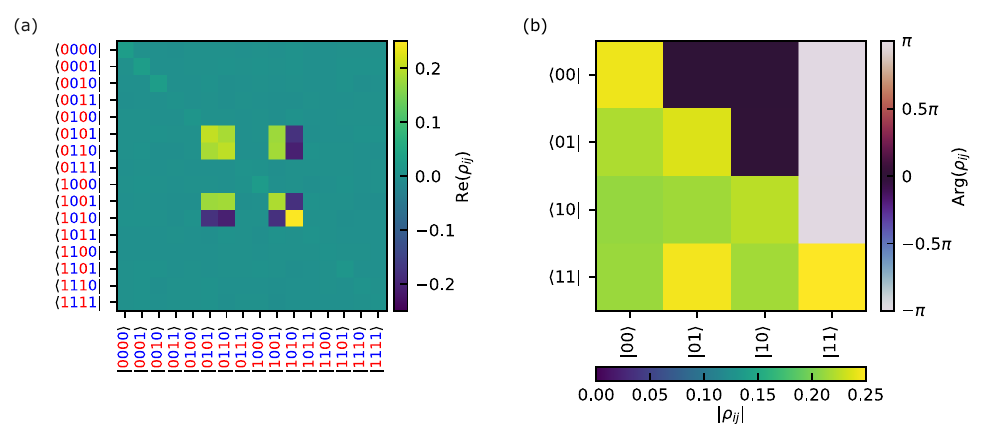}
\caption{Reconstructed density matrix of the generated two-logical-qubit frequency-bin cluster state. (a)~Density matrix of the reconstructed state in the bra--ket notation. The blue (red) colored modes correspond to the $\ket{\omega_1}(\ket{\omega_2})$ modes. Each adjacent pair (i.e., at positions 0 and 1, 2 and 3, etc.) corresponds to two modes within the same time bin. (b)~Density matrix of the reconstructed state, projected onto the logical subspace spanned by $\{\ket{\omega_1},\ket{\omega_2}\}$. Absolute values of the matrix elements are plotted in the diagonal and the lower-left triangle, and the complex arguments are plotted in the upper-right triangle.}
    \label{fig:freq_bin_results}
\end{figure*}

As mentioned in Sec.~\ref {sec:method}, we use the frequency-bin encoding protocol to generate dual-rail cluster states. By applying quantum-state tomography to itinerant microwave photons, we reconstruct the density matrix of the generated state~\cite{eichler2011experimental,eichler2012observation}. Details of the tomography can be found in Appendix~\ref{app_sec:tomo}. 
In Fig.~\ref{fig:freq_bin_results}(a), we show the density matrix of the generated frequency-bin cluster state consisting of two logical qubits. The reconstructed density matrices of three- and four-logical-qubit states can be found in Appendix~\ref{app_sec:App_exp}. The fidelities of the generated two-, three- and four-logical-qubit states are $76.4\pm0.8\%$, $67.4\pm2.1\%$, and $57.0\pm2.8\%$, respectively. From the density matrix, we see that the main state occupation is within the logical subspace that is spanned by $\ket{\omega_1}$ and $\ket{\omega_2}$ (corresponding to $\ket{01}$ and $\ket{10}$), and that there is also some occupation in $\ket{00}$ components. This indicates that there is photon loss during the state generation process, which is mainly induced by the qubit decay. However, following the method used in Refs.~\citenum{kurpiers2019quantum}~and~\citenum{ilves2020demand}, when the reconstructed density matrix is restricted to the logical subspace spanned by $\{\ket{\omega_1},\ket{\omega_2}\}$, effectively removing the influence of the losses, the fidelities of the corresponding states become $90.9\pm1.0\%$, $76.8\pm2.5\%$, and $67.8\pm3.4\%$, respectively. The corresponding results are shown in Fig.~\ref{fig:freq_bin_results}(b) and Appendix~\ref{app_sec:App_exp}. The remaining infidelity is caused by qubit decoherence during the generation sequence and the mode overlap. Details can be found in Appendix~\ref{app_sec:Theo_simu}.

\begin{figure}
    \centering
\includegraphics{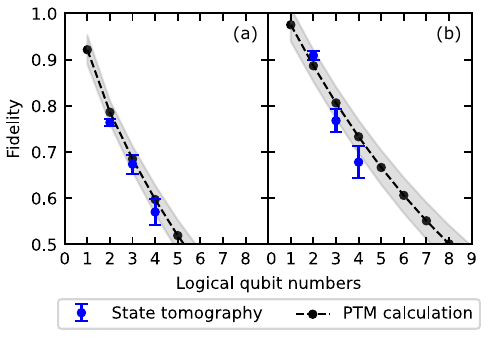}
    \caption{Fidelity of the generated photonic states: (a)~without and (b)~with photon loss correction, respectively. The blue dots are the fidelity of generated photonic states, which is reconstructed via state tomography. The black dots and dashed line represent the fidelities calculated based on the experimentally measured Pauli transfer matrix~(PTM). The gray colored areas present their standard deviations.}
    \label{fig:fidelity_PTM}
\end{figure}

We also apply process tomography for a single photon-emission process and find the process fidelity to be $86.7\pm0.7\%$. The details of the process tomography can be found in Appendix~\ref{subsec:method_process_tomo}. The results are close to the limit of this sample, which is $86.3\pm0.5\%$. Details of the calculation on the coherence limit can be found in Appendix~\ref{app_sec:Theo_simu}. From the results, we obtain the Pauli transfer matrix (PTM) of this photon emission process~\cite{besse2020realizing}. As shown in Fig.~\ref{fig:fidelity_PTM}, the fidelities of the generated states agree with the numerical simulation using the PTM. This indicates that we can assume the experimental photon-emission process in each emission round to be identical. Due to the large computational resource requirements for numerical simulation with more modes, we use an exponential function to estimate the fidelity of states with more modes. The validity of this estimation is supported by the properties of the generated state represented as an matrix product operator (MPO)~\cite{baumgratz2014efficient}.

Using the PTM, we can estimate the fidelities of states with more modes without carrying out quantum state tomography. From the PTM results shown in Fig.~\ref{fig:fidelity_PTM}, we find that for the generated photonic states, the fidelity can remain above 50$\%$ for up to five logical qubits. However, when we focus on the logical subspace, this number increases to as many as eight logical qubits, representing a significant improvement.

\subsection{Entanglement characterization in logical and physical modes}
Since we have reconstructed the density matrices of the dual-rail cluster states, we can now study the entanglement between different modes within the photonic state. Here, we use localizable entanglement~(LE) to evaluate how far the entanglement can be maintained in a cluster state. The LE can be calculated by applying projection measurements, except for two desired modes, and calculating the entanglement between them~\cite{verstraete2004entanglement,popp2005localizable}. The details of the LE calculation can be found in Appendix~\ref{app_sec:LE}. 

Fig.~\ref{fig:fig_LE}(a) shows the LE between different modes in the generated four-logical-qubit state. The results for two- and three-logical-qubit states can be found in Appendix~\ref{app_sec:App_exp}. The lower-left part of the figure shows the LE between the physical modes, while the upper-right part shows the LE between the logical qubits in the frequency-bin logical subspace after loss correction. From Fig.~\ref{fig:fig_LE}(a), we observe that the LE between two logical qubits is larger than the average LE between any pair of physical modes belonging to different logical qubits. Here, each logical qubit consists of two physical modes, so the comparison involves averaging over the four physical-mode pairs connecting the two logical qubits. It shows that the entanglement between different modes is protected by the logical encoding protocol. Figure~\ref{fig:fig_LE}(b)[(c)] shows the LE between the first photon (logical) mode and all other photon (logical) modes. We observe that, for the entanglement between the 0th mode and the other two modes which form the same logical qubit, there exists little difference. This is consistent with the graph representation of the state shown in Fig.~\ref{fig:scheme}(f). Each logical qubit is comprised of a pair of adjacent physical modes, leading to identical logical distances within each pair. As a result, the LE values exhibit plateau-like behavior across every two adjacent physical modes (e.g., the 2nd--3rd modes and the 4th--5th modes), reflecting their shared logical distance to the 0th mode.

In Figs.~\ref{fig:fig_LE}(b) and (c), we also plot the LE calculated from the measured PTM of each photon generation process. We find that it is consistent with the entanglement calculated from the reconstructed measured state. Also, based on this PTM, we can calculate how far the entanglement can be maintained within the graph state. Setting a threshold of LE at $0.05$ (10$\%$ of the maximally entangled value of an ideal state, 0.5), we find that the entanglement between physical modes remains up to 7 logical qubits. In contrast, in the logical subspace, the entanglement between logical qubits can be maintained up to 11 logical qubits. This indicates that frequency-bin dual-rail encoding can protect entanglement within the state. 
\begin{figure}
    \centering
\includegraphics{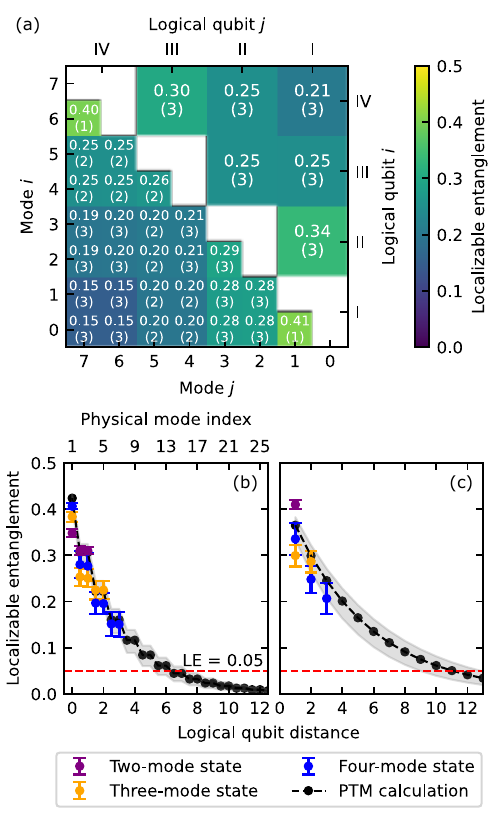}
    \caption{Localizable entanglement~(LE) and its standard deviation between different modes in the generated states. (a)~LE in a four-logical-qubit state. The lower-left part of the figure shows the LE between the physical modes, while the upper-right part corresponds to the LE between the corresponding logical qubits (indicated as I, II, III, and IV). The number in parentheses indicates the one-standard-deviation statistical uncertainty, referring to the uncertainty in the last digit of the LE value. (b)~LE between the 0th and other physical modes in two-, three-, and four-mode states. The bottom and top horizontal axes represent the physical and logical qubit distances, respectively, between the 0th mode and the target mode. The first data point corresponds to zero logical distance. (c)~LE between the logical qubit I and the other logical qubits in two-, three-, and four-mode states. In~(b) and~(c), the black dots and dashed line represent the LE calculated based on the experimentally measured Pauli transfer matrix (PTM). The gray colored areas present their standard deviations.}
    \label{fig:fig_LE}
\end{figure}

\subsection{Performance Comparison\label{sub_sec:exp_performance_comp}}
Finally, to benchmark the advantage of the frequency-bin encoding, we performed a numerical simulation of a conventional single-photon encoding scheme using identical device parameters. The results show that frequency-bin encoding maintains cluster-state fidelity above 50 $\%$ for up to eight logical qubits (8 vs 7) and exhibits a longer localizable entanglement length (12 vs 8), confirming its superior robustness under the same experimental constraints (see Appendix~\ref{app_sec:Theo_simu}). 
\section{CONCLUSION and discussion \label{sec:discussion}}
In this paper, we used a superconducting qubit to generate a single microwave photon state in the superposition of a pair of photon modes that have different frequencies. The generated photons can maintain entanglement with the superconducting qubit. We found that this photon generation process had a fidelity of $86.7\pm0.7\%$. Based on this process, we have generated dual-rail cluster states up to four logical qubits. The fidelities of these generated dual-rail cluster states were all above $50\%$. With error detection based on frequency-bin encoding, the fidelities were increased to larger than $60\%$, indicating that frequency-bin encoding can protect the cluster state. Based on the process tomography measurement results, we estimated that entanglement persists across up to 7 logical qubit (14 physical modes), and even up to 11 logical qubit when restricted to the frequency-bin-encoded logical subspace. We also compared our frequency-bin encoding protocol with conventional single-photon encoding simulated on the same device and achieved higher fidelity and preserved entanglement over a larger number of modes. These results demonstrate that frequency-bin encoding offers enhanced protection of entanglement and improved state fidelity, making it a promising approach for scalable photonic quantum information processing. 

An important consideration in our implementation is the spectral overlap between two frequency channels, which directly impacts their orthogonality and hence the validity of the dual-rail encoding. In our experiment, we chose a mode overlap of approximately 3$\%$ between two 1-$\mu$s-long photon pulses separated by 32~MHz. This overlap level strikes a balance between ensuring sufficient orthogonality for encoding logical qubits and maintaining compatibility with the detection bandwidth of our JPA used for quantum state tomography. At this overlap level, fidelity is primarily limited by qubit energy decay and decoherence rather than mode indistinguishability. This constraint also limits the possibility of reducing the spectral overlap simply by extending the photon pulse duration, since longer pulses would suffer more strongly from qubit energy decay and decoherence. Our choice thus reflects a practical trade-off, optimized for the constraints of our hardware. 
Furthermore, while the Purcell filter protects the qubit lifetime, the resonator's effective linewidth constrains the maximum frequency separation. To remove this constraint, the multiple resonator modes induced by the resonator--filter system can be engineered to obtain a wider spectral range for the resonator--transmission-line coupling~\cite{miyamura2025,Miyamura2025SQA}. This extends the tunable frequency range, resulting in a larger possible separation between frequency channels. 

Although we implemented erasure detection via post-selection on the reconstructed density matrix in this work, real-time photon loss detection on frequency-bin dual-rail states is also feasible. Inspired by the approach in Ref.~\citenum{reuer2022realization}, by using an additional cQED system, we can coherently map two co-propagating photon modes onto the two energy levels of a superconducting qubit via a controlled absorption process. By distinguishing the qubit state whether it's in the subspace spanned by the target levels or not~\cite{jerger_realization_2016}, photon loss events can be selectively detected while maintaining the coherence in the subspace. This enables the implementation of real-time photon loss detection. 

In the present implementation, the generation of the frequency-bin cluster states relies on post-selection of the qubit in the $\ket{g}\pm\ket{e}$ basis. This post-selection is required because the qubit remains entangled with the emitted photons, which is essential for scalable photonic state generation and cannot be disentangled deterministically within the same driving sequence. Nevertheless, a fully deterministic termination of the sequence can be incorporated by replacing only the final photon-emission step with simultaneously driven second-order transitions such as $\ket{e0}$--$\ket{g1}$ together with $\ket{f0}$--$\ket{g1}$. Using this modified last emission step returns the qubit to its ground state while still emitting a photon into well-defined frequency modes, without altering the scalable generation steps preceding it. This approach, demonstrated in Refs.~\citenum{Miyamura2025SQA,yang2024deterministic}, would enable unconditional state generation, making the source more suitable for scalable and real-time applications.

Another important direction concerns improving the overall state fidelity toward practical applications. For the applications discussed in the introduction, such as measurement-based quantum computing and quantum repeaters, a fidelity above 90$\%$ is generally required to ensure fault-tolerant operation~\cite{raussendorf2006fault,azuma2015all}. In our current implementation, the main limitation arises from the finite qubit energy and coherence lifetimes, as the photon generation process has not yet reached the coherence-limited regime. Enhancing material quality and fabrication~\cite{bland2025millisecond,tuokkola2025methods,bal2024systematic}, along with suppression of quasiparticle~\cite{wang2014measurement,mcewen2024resisting,connolly2024coexistence} and two-level-system losses~\cite{colao2025mitigating,chen2024phonon,lisenfeld2023enhancing}, would help mitigate these limitations. In addition, optimized pulse shaping for photon generation~\cite{miyamura2025} can reduce leakage and further improve the process fidelity.
As shown in Appendix~\ref{app_sec:Theo_simu}, an ideal qubit with infinite lifetime would yield a process fidelity exceeding $97.2\%$, suggesting that the fidelities required for scalable applications are within reach of near-future devices.

Our frequency-bin scheme is compatible with many of the existing approaches for increasing the dimensionality of a photonic cluster state. For example, in the optical domain, many advances in cluster-state generation have occurred, typically using continuous-variable approaches. These approaches typically generate two-mode entanglement through the interference between two squeezed-light sources, and optical delay lines are used to scale up the dimension of the state~\cite{larsen2019deterministic,asavanant2019generation,roh2025generation}. In the microwave domain, recent works have demonstrated 2D cluster-state generation~\cite{ferreira2024deterministic,o2024deterministic}. Our protocol is compatible with the underlying principles of these previous works.

Looking ahead, our protocol could be extended by increasing the number of time bins to generate longer 1D cluster states. Furthermore, integrating multiple qubits or using qubit arrays could enable spatial multiplexing, opening the possibility of generating higher dimension cluster states. Such higher dimension structures are a key ingredient for realizing fault-tolerant measurement-based quantum computation~\cite{raussendorf2006fault,raussendorf2007topological}. The compatibility of frequency-bin dual-rail encoding with erasure detection and its robustness to loss make it particularly promising for scaling up quantum network protocols and photonic quantum processors in the microwave regime.

\begin{acknowledgments}
We thank S. Tamate, P.A. Spring, S. Watanabe, K. Kato, K. Yuki, S. Inoue, S. Kikura and R. Zainudin for fruitful discussions.
This work was supported by the Ministry of Education, Culture, Sports, Science and Technology (MEXT) Quantum Leap Flagship Program (Q-LEAP) (Grant No. JPMXS0118068682), and the JSPS Grant-in-Aid for Scientific
Research (KAKENHI) (Grant No. JP22H04937).

\end{acknowledgments}
\clearpage
\appendix

\section{Devices and experimental setup\label{app_sec:setup}}
The setup of our experiment is shown in Fig.~\ref{app_fig:setup}. 
We generate the qubit drive pulses by up-converting an intermediate frequency (IF) pulse generated by an arbitrary waveform generator (AWG) using an I/Q (in-phase and quadrature) mixer and a local oscillator. The required drive frequencies for the $\ket{g}$--$\ket{e}$, $\ket{e}$--$\ket{f}$, and $\ket{f}$--$\ket{h}$ transitions span a range of around $700$~MHz. Because this range is larger than the bandwidth of our AWG, $250$~MHz, we used two sets of AWGs, mixers, and local oscillators (LOs) to generate the three qubit drives. In addition, we used two more sets to generate the $\ket{f0}$--$\ket{g1}$ and $\ket{h0}$--$\ket{e1}$ pulses and another for the readout pulse. All the LOs, AWGs and ADC are phase locked by the same $10$-MHz rubidium clock to synchronize them. In order to keep the phase of the single-photon signal coherent over multiple measurements, the local oscillator frequencies of the these microwave sources need to satisfy the following equations~\cite{ilves2020demand}:
\begin{figure}[t]
    \centering
\includegraphics{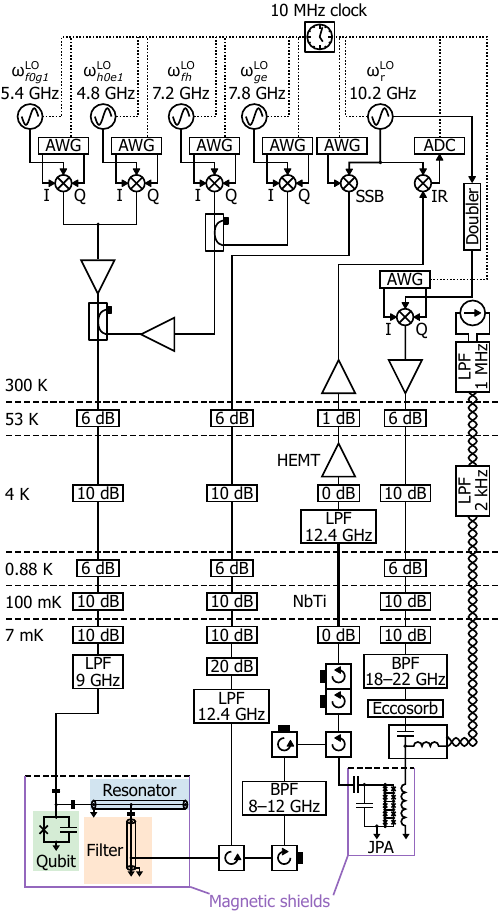}
    \caption{Measurement setup used in the experiment. AWG, arbitrary waveform generator; ADC, analog-to-digital converter; SSB, single-sideband mixer; IR, image reject mixer; LPF, low-pass filter; HEMT, high-electron-mobility transistor; BPF, band-pass filter; and JPA, Josephson parametric amplifier. LO, the local oscillator used for the microwave drives. From left to right, the one for the external $\ket{f0}$--$\ket{g1}$ drive, $\omega_{f0g1}^{\mathrm{LO}}$, the one for the external $\ket{h0}$--$\ket{e1}$ drive, $\omega_{h0e1}^{\mathrm{LO}}$, the one for the qubit $\ket{f}$--$\ket{h}$ drive, $\omega_{fh}^{\mathrm{LO}}$, the one for qubit $\ket{g}$--$\ket{e}$ and $\ket{e}$--$\ket{f}$ drives, $\omega_{ge}^{\mathrm{LO}}$, and the one for the qubit readout, photon measurement, and JPA pump, $\omega_{\mathrm{r}}^{\mathrm{LO}}$. All the LOs, AWGs and ADC are phase locked by the same $10$-MHz rubidium clock, shown as the dot line connected to the clock.}
    \label{app_fig:setup}
\end{figure}

\begin{equation}
    \begin{split}
        \omega_{f0g1}^{\mathrm{LO}}+\omega_{\mathrm{r}}^{\mathrm{LO}}-2\omega_{ge}^{\mathrm{LO}}&=0,\\
        \omega_{h0e1}^{\mathrm{LO}}+\omega_{\mathrm{r}}^{\mathrm{LO}}-\omega_{ge}^{\mathrm{LO}}-\omega_{fh}^{\mathrm{LO}}&=0,
    \end{split}
\end{equation}

The energy-relaxation and dephasing times of the qubit are listed in Table~\ref{tab:sys_parameter}. The relaxation times between the higher levels were extracted using the method described in Ref.~\citenum{pechal2016microwave}, where additional pulse sequences are used to cancel the effect of relaxation out of the relevant subspace. For the Ramsey dephasing times, the relaxation out of the two-level subspace was also considered in the fitting. The bare parameters for the resonator and the filter are also listed.

 \begin{table}[t]
    \caption{System parameters.}
    \centering
    \begin{ruledtabular}
    \begin{tabular}{lcr}
    Qubit $\ket{g}$--$\ket{e}$ frequency & $\omega_{ge}/2\pi$ (MHz) & 8021.8\\
    Qubit $\ket{e}$--$\ket{f}$ frequency & $\omega_{ef}/2\pi$ (MHz) & 7702.9\\
    Qubit $\ket{f}$--$\ket{h}$ frequency & $\omega_{fh}/2\pi$ (MHz) & 7347.6\\
    \hline
    $\ket{g}$--$\ket{e}$ energy-relaxation time & $T_1^{ge}$ ($ \mathrm{\mu s}$) & $32.6\pm5.0$ \\
    $\ket{g}$--$\ket{e}$ Ramsey dephasing time & $T_2^{ge}$ ($ \mathrm{\mu s}$) &$21.5\pm4.4$\\
    $\ket{g}$--$\ket{e}$ echo dephasing time & $T_{2e}^{ge}$ ($ \mathrm{\mu s}$) &$40.1\pm4.3$\\
    $\ket{e}$--$\ket{f}$ energy-relaxation time & $T_1^{ef}$ ($ \mathrm{\mu s}$) &$23.0\pm 1.6$\\
    $\ket{e}$--$\ket{f}$ Ramsey dephasing time & $T_2^{ef}$ ($ \mathrm{\mu s}$) &$10.3\pm1.4$\\
    $\ket{f}$--$\ket{h}$ energy-relaxation time & $T_1^{fh}$ ($ \mathrm{\mu s}$) &$11.3\pm2.2$\\
    $\ket{f}$--$\ket{h}$ Ramsey dephasing time & $T_2^{fh}$ ($ \mathrm{\mu s}$) &$4.8\pm0.9$\\
    \hline
    Resonator dressed frequency, $\ket{g}$ & $\omega_\mathrm{r}^{g}/2\pi$ (MHz) &10299.5\\
    Dispersive shift & $2\chi/2\pi$ (MHz) & 4.1\\
    Resonator bare frequency & $\omega_\mathrm{r}/2\pi$ (MHz) &  10286.6\\
    Filter frequency & $\omega_\mathrm{f}/2\pi$ (MHz) & 10273.5\\
    Filter external coupling &$\kappa_\mathrm{f}/2\pi$ (MHz)&449.3\\
    Resonator--filter coupling strength &$J/2\pi$ (MHz) &94.6\\
    Resonator--qubit coupling strength &$g/2\pi$ (MHz)  &192.9\\
\end{tabular}
    \end{ruledtabular}
    \label{tab:sys_parameter}
\end{table}

To get the system parameters shown in Table~\ref{tab:sys_parameter}, we measure the spectra of the resonator when the qubit state is in the ground and excited states, $S_{11}^g$ and $S_{11}^e$. We determine the parameters by fitting the measured $S_{11}^g/S_{11}^e$ and qubit frequencies with calculated $S_{11}^{g(c)}/S_{11}^{e(c)}$ and qubit frequencies from the following Hamiltonian:
\begin{equation}
\begin{split}
    \hat{H}_{\mathrm{sys}}/\hbar&=\omega_\mathrm{q} \hat{b}^\dagger\hat{b}+\frac{\alpha}{2}\hat{b}^\dagger\hat{b}^\dagger\hat{b}\hat{b}+\frac{\alpha_\mathrm{h}}{6}\hat{b}^\dagger\hat{b}^\dagger\hat{b}^\dagger\hat{b}\hat{b}\hat{b}\\    &+\omega_\mathrm{r}\hat{a}^\dagger\hat{a}+\omega_\mathrm{f}\hat{f}^\dagger\hat{f}\\
    &+g(\hat{a}\hat{b}^\dagger+\hat{a}^\dagger\hat{b})+ J(\hat{a}\hat{f}^\dagger+\hat{a}^\dagger\hat{f}).
\end{split}
\label{app_eq:Hamil}
\end{equation}
Here, $\alpha=\omega_{ef}-\omega_{ge}$ is the anharmonicity between qubit $\ket{e}$--$\ket{f}$ and $\ket{g}$--$\ket{e}$ transitions, and $\alpha_{\mathrm{h}}=\omega_{fh}-\omega_{ef}-2\alpha$ is the higher order nonlinearity for the $\ket{h}$ state, and $\hat{a},\hat{b},\hat{f}$ represent the resonator mode, the qubit mode and the filter mode respectively. The spectra $S_{11}^{g(c)}$ and $S_{11}^{e(c)}$ can be obtained by calculating the correlation function $\langle\hat{f}(t)\hat{f}^\dagger(0)\rangle_{\ket{g}}$ and $\langle\hat{f}(t)\hat{f}^\dagger(0)\rangle_{\ket{e}}$ based on the Wiener–Khinchin theorem~\cite{cohen1998generalization}. The fitted parameters and the Hamiltonian are used in the simulations in Appendix~\ref{app_sec:Theo_simu}.

\section{Photon calibration\label{app_sec:cali}}
\subsection{Raman-process calibration}

To implement the two Raman processes, the $\ket{f0}$--$\ket{g1}$ and $\ket{h0}$--$\ket{e1}$ transitions, it is essential to calibrate their ac Stark shifts accurately. In our experiment, the drive strengths are controlled via the voltage amplitudes applied to the arbitrary waveform generator (AWG), but the exact relation between this voltage and the physical drive strength $\Omega_\mathrm{d}$ is not known a priori. We therefore perform Stark shift measurements to extract this relationship.

We first measure the Stark shift of each Raman transition individually by applying only one drive at a time. This allows us to isolate the effect of each drive and fit the observed frequency shift as a function of applied voltage and drive detuning $\delta_\mathrm{q} = \omega_\mathrm{q} - \omega_\mathrm{d}$. Using the second-order perturbation theory, the ac Stark shift $\delta$ under a detuned drive $\Omega_\mathrm{d}$ can be modeled as:
\begin{equation}
    \begin{split}
    \delta_{f0g1}(\Omega_{\mathrm{d}},\delta_{\mathrm{q}})&=\frac{\alpha(2\delta_{\mathrm{q}}+\alpha)}{2\delta_{\mathrm{q}}(\delta_{\mathrm{q}}+\alpha)(\delta_{\mathrm{q}}+2\alpha)}\Omega_{\mathrm{d}}^2,\\
    \delta_{h0e1}(\Omega_{\mathrm{d}},\delta_{\mathrm{q}})&=\frac{\alpha(2\delta_{\mathrm{q}}+3\alpha)(\delta_{\mathrm{q}}-\alpha)}{2\delta_{\mathrm{q}}(\delta_{\mathrm{q}}+\alpha)(\delta_{\mathrm{q}}+2\alpha)(\delta_{\mathrm{q}}+3\alpha)}\Omega_{\mathrm{d}}^2.
    \end{split}
    \label{app-eq:stark_shift}
\end{equation}

\begin{figure}
    \centering
\includegraphics{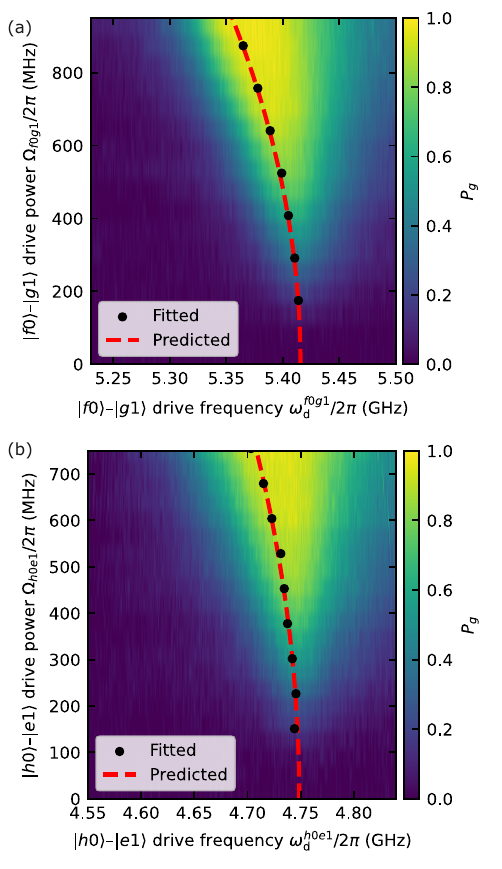}
    \caption{Ac Stark shifts of two Raman processes under individual drives and their fits: (a)~the $\ket{f0}$--$\ket{g1}$ transition, and (b)~the $\ket{h0}$--$\ket{e1}$ transition. The black dots are the fitted resonance frequencies based on a Fano-resonance model to account for slight asymmetry in the measured spectra, while the red dashed lines are the predicted frequencies based on Eq.~\eqref{app-eq:stark_shift}.}
    \label{app_fig:stark_single}
\end{figure}

Figure~\ref{app_fig:stark_single}(a) shows the Stark shift of the $\ket{f0}$--$\ket{g1}$ transition under varying voltage and detuning of the $\ket{f0}$--$\ket{g1}$ drive, corresponding to $\delta_{f0g1}(\Omega_{f0g1},\omega_{\mathrm{q}}-\omega_\mathrm{d}^{f0g1})$.  
Similarly, Fig.~\ref{app_fig:stark_single}(b) shows the shift of the $\ket{h0}$--$\ket{e1}$ transition under the $\ket{h0}$--$\ket{e1}$ drive, corresponding to $\delta_{h0e1}(\Omega_{h0e1},\omega_{\mathrm{q}}-\omega_\mathrm{d}^{h0e1})$. By fitting these curves to Eq.~\eqref{app-eq:stark_shift}, we extract the proportionality between the AWG voltage and the physical drive strength $\Omega_\mathrm{d}$.

To generate the dual-rail cluster states, we apply $\ket{h0}$--$\ket{e1}$ and $\ket{f0}$--$\ket{g1}$ drives simultaneously. Compared with the $\ket{f0}$--$\ket{g1}$ transition, the $\ket{h0}$--$\ket{e1}$ transition is more susceptible to additional Stark shifts induced by another drive. Therefore, we focus on calibrating the $\ket{h0}$--$\ket{e1}$ transition under simultaneous drives.

\begin{figure}
    \centering
\includegraphics{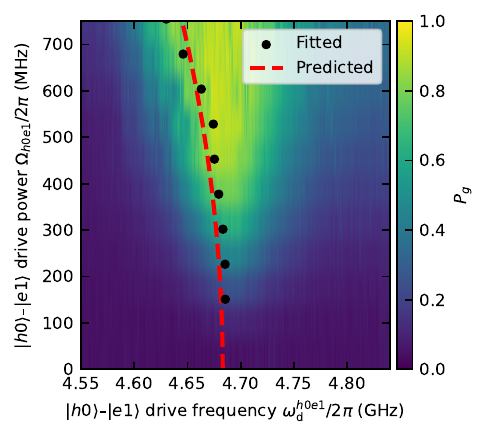}
    \caption{Ac Stark shift of the $\ket{h0}$--$\ket{e1}$ transition under an additional $\ket{f0}$--$\ket{g1}$ drive with a power of $\Omega_{f0g1}/2\pi=699$~MHz at a frequency of $\omega_\mathrm{d}^{f0g1}/2\pi=5400$~MHz. The black dots are the fitted resonance frequencies based on a Fano-resonance model to account for slight asymmetry in the measured spectra, while the red dashed line is the predicted frequencies based on Eq.~\eqref{app-eq:stark_shift}.}
    \label{app_fig:stark_double}
\end{figure}

Here, we apply the $\ket{f0}$--$\ket{g1}$ drive at selected frequencies and strengths (chosen based on Fig.~\ref{app_fig:stark_single}(a)) and scan the $\ket{h0}$--$\ket{e1}$ drive to measure its resonance frequency shift. Figure~\ref{app_fig:stark_double} shows the result for a specific case with $V_{f0g1} = 0.6$~V and $\omega_\mathrm{d}^{f0g1}/2\pi=5400$~MHz. The observed shift is larger than in the single-drive case due to the additional Stark contribution from the $\ket{f0}$--$\ket{g1}$ drive:
\begin{equation}
    \delta_{h0e1}(\Omega_{h0e1},\omega_{\mathrm{q}}-\omega_\mathrm{d}^{h0e1})+\delta_{h0e1}(\Omega_{f0g1},\omega_{\mathrm{q}}-\omega_\mathrm{d}^{f0g1}).
\end{equation}

From these calibrations, we determine the drive strengths and frequencies used in our experiment:
\begin{itemize}
    \item $V_{f0g1} = 0.6$~V → $\Omega_{f0g1}/2\pi = 699$~MHz  
    \item $V_{h0e1} = 0.7$~V → $\Omega_{h0e1}/2\pi = 528$~MHz  
    \item $\omega_\mathrm{d}^{f0g1}/2\pi = 5405$~MHz, $\omega_\mathrm{d}^{h0e1}/2\pi = 4665$~MHz
\end{itemize}
These values are used in the numerical simulations presented in Appendix~\ref{app_sec:Theo_simu}.

\subsection{Photon-generation calibration\label{app_sub_sec:photon-generation-cali}}
\begin{figure}
    \centering
\includegraphics{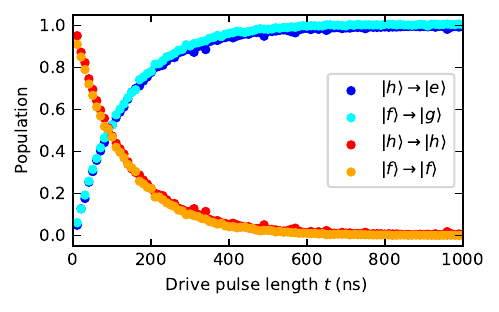}
    \caption{Qubit population as a function of the drive pulse length $t$ under different initially prepared states. For the qubit prepared in the $\ket{h}$ state, we show the measured populations of the $\ket{h}$ state (red dots) and the $\ket{e}$ state (blue dots) after applying the $\ket{h0}$--$\ket{e1}$ and $\ket{f0}$--$\ket{g1}$ drives simultaneously. For the qubit prepared in the $\ket{f}$ state, we show the measured populations of the $\ket{f}$ state (orange dots) and the $\ket{g}$ state (cyan dots) after applying the $\ket{h0}$--$\ket{e1}$ and $\ket{f0}$--$\ket{g1}$ drives simultaneously.}
    \label{app-fig:time_pop}
\end{figure}

\textbf{Drive length calibration} To calibrate the photon generation process, we prepare the qubit in the second (third) excited state, $\ket{f}$~($\ket{h}$), then apply two drives ($\ket{f0}$--$\ket{g1}$ and $\ket{h0}$--$\ket{e1}$) simultaneously. By varying the drive pulse length, we measure the qubit population in each state, as shown in Fig.~\ref{app-fig:time_pop}. The photon emission rate of each transition is obtained from an exponential decay, resulting in a decay time of $1/\Gamma_{\ket{f}}^{f0g1}=0.131~\mathrm{\,\mu s}$ and $1/\Gamma_{\ket{h}}^{h0e1}=0.135~\mathrm{\,\mu s}$. Thus, we choose a pulse length of $1~\mathrm{\,\mu s}$ to ensure the completeness of the emitted photon.

\textbf{Photon frequency calibration} To get the spectra of two photon modes, we measure the photon waveform in the time domain. For the waveform of photon mode $\ket{\omega_1}$, corresponding to the $\ket{f0}$--$\ket{g1}$ transition, one can simply prepare the qubit state at $(\ket{g}+\ket{f})/\sqrt{2}$ and apply the external drives to get the $(\ket{0}+\ket{1})/\sqrt{2}$ photon. In practice, we not only measure the $(\ket{0}+\ket{1})/\sqrt{2}$ photon, but also measure the $(\ket{0}-\ket{1})/\sqrt{2}$ by preparing the qubit state at $(\ket{g}-\ket{f})/\sqrt{2}$. Taking the difference of these two measured waveforms with opposite signs, we cancel the background noise while preserving the temporal shape of the signals~\cite{sunada2024efficient,miyamura2025}. For the waveform of photon mode $\ket{\omega_2}$, we also get the photon waveform shape by preparing the qubit state the ($\ket{e}\pm\ket{h})/\sqrt{2}$ and get the waveform by taking their difference. After getting the waveforms in the time domain, we use the Fourier transformation to get their spectra, as shown in Fig.~\ref{fig:spectra}. And for the spectra of $(\ket{g}+\ket{e}+\ket{f}+\ket{h})/2$, it was also obtained by taking the difference with the state prepared at $(\ket{g}+\ket{e}-\ket{f}-\ket{h})/2$.
While for the simplicity, we still use the notation of $(\ket{g}+\ket{f})/\sqrt{2}$, $(\ket{e}+\ket{h})/\sqrt{2}$ and$(\ket{g}+\ket{e}-\ket{f}-\ket{h})/2$ in the main text. All spectra in Fig.~\ref{fig:spectra} are normalized by the same constant factor for visual comparison, such that the maximum amplitude corresponds to unity. This normalization does not affect any physical quantity.

\begin{figure}
    \centering
\includegraphics{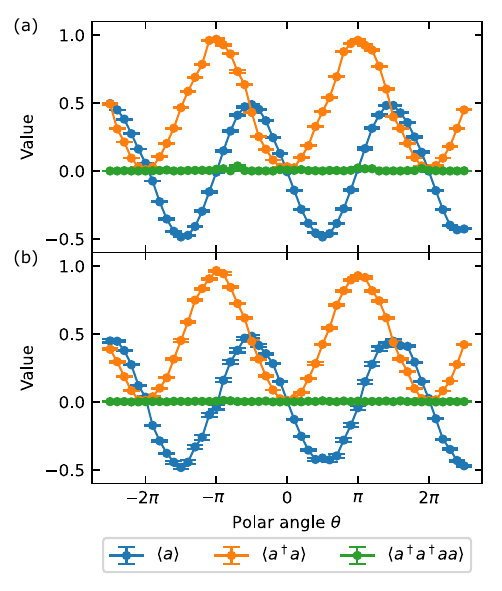}
    \caption{Moments of an emitted photon as a function of the preparation angle $\theta$ of the source qubit. (a)~Photon emitted from the $\ket{f0}$--$\ket{g1}$ transition. (b)~Photon emitted from the $\ket{h0}$--$\ket{e1}$ transition.}
    \label{app-fig:single_photons}
\end{figure}

\textbf{Single photon property calibration} We characterize the emitted single photon in each process using the tomography method described in Appendix~\ref{subsec:method_tomo}. We prepare the qubit in a superposition state $\cos\frac{\theta}{2}\ket{g}+\sin\frac{\theta}{2}\ket{f}$ ($\cos\frac{\theta}{2}\ket{e}+\sin\frac{\theta}{2}\ket{h}$) for the $\ket{f0}$--$\ket{g1}$ ($\ket{h0}$--$\ket{e1}$) photon, and then apply both external drives to emit the photon. By rotating the polar angle $\theta$, we measure the moments of the emitted photon. The results are shown in Fig.~\ref{app-fig:single_photons}. 

The measured fourth-order moment $\braket{\hat{a}^\dagger\hat{a}^\dagger\hat{a}\hat{a}}$ is always close to 0, indicating a single-photon process. Thus, in the following state reconstruction, we choose a truncation into two dimensions, $\{\ket{0},\ket{1}\}$, for each mode. Also, based on these results, we obtain the scaling factor and measurement efficiency of each photon mode~\cite{eichler2012observation}. The measurement efficiency is $29.4\%$ for the $\ket{f0}$--$\ket{g1}$ mode, and $22.2\%$ for the $\ket{h0}$--$\ket{e1}$ mode.
 \section{Details of the state generation\label{app_sec:graph_repre_top}}
\subsection{The equivalence of pulse sequence\label{app_sec:generation_seq}}

As discussed at the end of Sec.~\ref{sec:method}, the pulse sequence shown in Fig.~\ref{fig:scheme}(e) provides an experimental realization of the conceptual circuit in Fig.~\ref{fig:scheme}(d). Here, we clarify the correspondence between them.

A conditional photon-emission process controlled by the transmon state can be regarded as a CNOT gate between the qubit and the corresponding photonic mode~\cite{sunada2024efficient,o2024deterministic,besse2020realizing}. In the experimental sequence of Fig.~\ref{fig:scheme}(e), the four $\pi$ pulses, ${\pi_{ef}, \pi_{ge}, \pi_{fh}, \pi_{ef}}$ [labeled (i)–(iv) in Fig.~\ref{fig:scheme}(e)] successively map the qubit population from the $\{\ket{g}, \ket{e}\}$ subspace to the $\{\ket{f}, \ket{h}\}$ subspace. Consequently, the subsequent $\ket{f0}$--$\ket{g1}$ and $\ket{h0}$--$\ket{e1}$ transitions are conditioned on the original $\ket{g}$ and $\ket{e}$ components, respectively.

The photon emission associated with the $\ket{h0}$--$\ket{e1}$ transition [red-colored photons with even indices in Fig.~\ref{fig:scheme}(d)] therefore corresponds to the first CNOT operation in the circuit.
For the $\ket{f0}$--$\ket{g1}$ transition [blue-colored photons with odd indices], the process starts and ends in the $\ket{g}$ state. This is equivalent to a NOT–CNOT–NOT sequence, effectively reproducing the second CNOT gate in Fig.~\ref{fig:scheme}(d) with surrounding X operations on the qubit.

The Hadamard gate in the circuit acts within the $\{\ket{g}, \ket{e}\}$ subspace to create a superposition of the two qubit states, $\ket{g}$ and $\ket{e}$
After mapping the population to the ${\ket{f}, \ket{h}}$ manifold, this operation is implemented experimentally by the $\pi_{fh}/2$ pulse [labeled (vi) in Fig.~\ref{fig:scheme}(e)], which plays the equivalent role of the Hadamard in the effective two-level subspace.

In this way, the overall pulse sequence in Fig.~\ref{fig:scheme}(e) faithfully reproduces the logical structure of the dual-rail cluster-state generation circuit in Fig.~\ref{fig:scheme}(d).

\subsection{Graph representation of the generated state\label{app_sec:graph_repre}}
As we mentioned in Fig.~\ref{fig:scheme}(f), the generated frequency-bin-encoded dual-rail cluster states can be represented as a comb-shaped graph state. Here, we show that they are locally equivalent to each other by applying local operators. For simplicity, we show the case of the two-logical-qubit state.

Based on the definition of a graph state~\cite{hein2004multiparty}, a four-mode comb-shaped graph state, which only contains the 0th, 1st, 2nd and 3rd mode in Fig.~\ref{fig:scheme}(f), is given by
\begin{equation}
\ket{\psi_{\mathrm{comb}}}=U_{\mathrm{CZ}}^{0,1}U_{\mathrm{CZ}}^{1,3}U_{\mathrm{CZ}}^{2,3}\ket{+}\ket{+}\ket{+}\ket{+}.
\end{equation}
Here, $U_{\mathrm{CZ}}^{a,b}$ is the controlled-Z gate between mode $a$ and mode $b$, and $\ket{\pm}=(\ket{0}\pm\ket{1})/\sqrt{2}$. This state can be rewritten as
\begin{equation}
\begin{split}
    \ket{\psi_{\mathrm{comb}}}=&\frac{1}{2}\bigl(\ket{+}\ket{0}\ket{+}\ket{0}+\ket{+}\ket{0}\ket{-}\ket{1}\\
    &+\ket{-}\ket{1}\ket{+}\ket{0}-\ket{-}\ket{1}\ket{-}\ket{1}\bigr).
\end{split}
\label{app_eq:comb}
\end{equation}
The above state can be locally transformed into the two-logical-qubit cluster state,
\begin{equation}
\begin{split}
    \ket{\psi_{\mathrm{L}}}=&\frac{1}{2}\left(\ket{\omega_1}\ket{\omega_1}+\ket{\omega_1}\ket{\omega_2}+\ket{\omega_2}\ket{\omega_1}-\ket{\omega_2}\ket{\omega_2}\right)\\
    =&\frac{1}{2}\left(\ket{0101}+\ket{0110}+\ket{1001}-\ket{1010}\right),
\end{split}
\label{app_eq:logical_cluster}
\end{equation}
by applying the Hadamard transform to the first and third modes. Therefore, frequency-bin-encoded dual-rail cluster states can be represented as comb-shaped graph states. We use this transformation when calculating the localizable entanglement between physical modes.

On the other hand, if the vertical edge of entanglement is on the even frequency channels instead of the odd ones shown in Fig.~\ref{fig:scheme}(f), it is still a graph representation of the genretaed state. Again, a four-mode comb-shaped graph state in this representation would be:
\begin{equation}
\begin{split}
\ket{\psi_{\mathrm{comb}}'}=&U_{\mathrm{CZ}}^{0,1}U_{\mathrm{CZ}}^{0,2}U_{\mathrm{CZ}}^{2,3}\ket{+}\ket{+}\ket{+}\ket{+}\\
    =&\frac{1}{2}\bigl(\ket{0}\ket{+}\ket{0}\ket{+}+\ket{1}\ket{-}\ket{0}\ket{+}\\
    &+\ket{0}\ket{+}\ket{1}\ket{-}-\ket{1}\ket{-}\ket{1}\ket{-}\bigr).
\end{split}
\end{equation}
By applying the Hadamard transform to the second and fourth modes, the state is transformed into Eq.~\eqref{app_eq:logical_cluster} again. Since both representations are equivalent based on the local operators, they have the same underlying entanglement structure. In this paper, we use the represenatation shown in Fig.~\ref{fig:scheme}(f).

 \section{Tomography of generated photonic states\label{app_sec:tomo}}
Based on the our measurement setup, we can apply a heterodyne-based state tomography to the generated photonic states~\cite{eichler2011experimental,eichler2012observation}.

\subsection{Itinerant photonic-state tomography \label{subsec:method_tomo}}
For each photonic mode, a complex amplitude $S_m=I_m+iQ_m$ can be measured. This measured complex amplitude contains both the photon signal and the noise in the detection chain, which is written as $\hat{S}=\hat{a}+\hat{h}^\dagger$~\cite{eichler2011experimental,eichler2012observation}. Here $\hat{a}$ is the annihilation operator of the signal photon and $\hat{h}^\dagger$ is the creation operator of the noise in the detection chain. In order to remove the noise from the measured signal, the complex amplitude of the vacuum-state signal, $S_{\rm{vac}}$, is necessary, which is $\hat{S}_{\rm{vac}}=\hat{h}^\dagger$. Therefore, the moment $\langle(\hat{S}^\dagger)^m\hat{S}^n\rangle$ satisfies the following equation:
\begin{equation}
\langle(\hat{S}^\dagger)^m\hat{S}^n\rangle=\sum_{i,j=0}^{m,n}\binom{n}{j}\binom{m}{i}\langle(\hat{a}^\dagger)^i\hat{a}^j\rangle\langle\hat{h}^{m-i}(\hat{h}^\dagger)^{n-j}\rangle,
\label{eq:single_mode_moment}
\end{equation}
where $\langle\hat{h}^{m-i}(\hat{h}^\dagger)^{n-j}\rangle$ is obtained from $\langle(\hat{S}_{\rm{vac}}^\dagger)^{m-i}\hat{S}_{\rm{vac}}^{n-j}\rangle$. Here we suppose that the noise mode has no correlation with the signal mode. By solving these equations, the moments $\langle(\hat{a}^\dagger)^m\hat{a}^n\rangle$ of the signal photon state can be obtained.

With the measured moments $\langle(\hat{a}^\dagger)^m\hat{a}^n\rangle$ and their corresponding standard deviations $\delta_{m,n}$, we use a maximum likelihood method to figure out the most likely density matrix $\hat{\rho}$ of the state~\cite{cvxpy}. The standard deviations $\delta_{m,n}$ are obtained through a nonparametric resampling process: we repeatedly resample the raw experimental data, calculate the moment $\langle(\hat{a}^\dagger)^m\hat{a}^n\rangle$ for each resampled dataset, and evaluate the standard deviation across these resampled estimates. The target function can be
\begin{equation}
    L=-\sum_{m,n}\frac{1}{\delta_{m,n}^2}\left|\langle(\hat{a}^\dagger)^m\hat{a}^n\rangle-\mathrm{Tr}\left(\hat{\rho}(\hat{a}^\dagger)^m\hat{a}^n\right)\right|^2,
\end{equation}
with constraints $\hat{\rho}\ge 0$ and $\mathrm{Tr}(\hat{\rho})=1$.

We can also use this method for multi-mode photonic states. Here, we take a two-photon-mode state as an example. For a two-photon-mode state, its moment satisfies the following equation:
\begin{equation}
\begin{split}
\langle(\hat{S}_1^\dagger)^m\hat{S}_1^n&(\hat{S}_2^\dagger)^p\hat{S}_2^q\rangle=\\
\sum_{i,j,k,l=0}^{m,n,p,q}&\binom{n}{j}\binom{m}{i}\binom{p}{k}\binom{q}{l}\langle(\hat{a}^\dagger)^i\hat{a}^j(\hat{b}^\dagger)^k\hat{b}^l\rangle\\
&\times\langle\hat{h}_1^{m-i}(\hat{h}_1^\dagger)^{n-j}\hat{h}_2^{p-k}(\hat{h}_2^\dagger)^{q-l}\rangle,
\end{split}
\label{eq:bi_mode_moment}
\end{equation}
and the corresponding target function is
\begin{equation}
\begin{split}
    L=&-\sum_{m,n,p,q}\frac{1}{\delta_{m,n,p,q}^2}\\
    &\times\left|\langle(\hat{a}^\dagger)^m\hat{a}^n(\hat{b}^\dagger)^p\hat{b}^q\rangle-\mathrm{Tr}\left(\hat{\rho}(\hat{a}^\dagger)^m\hat{a}^n(\hat{b}^\dagger)^p\hat{b}^q\right)\right|^2.
\end{split}    
\label{eq:multi_mode_reconstruct}
\end{equation}

For this reconstruction method, a sufficient amount of data sampling is necessary. A simple estimation of the required amount of data can be given by the following formula~\cite{da2010schemes,eichler2013experimental}:

\begin{equation}
    N=(1+N_0)^{o}.
    \label{eq:data_noise_relation}
\end{equation}
Here, $N_0$ is the effective noise photon number in the detection chain and $N$ is the least sampling number required to get a moment with order $o$. For example, for the moment $\langle(\hat{a}^\dagger)^m\hat{a}^n\rangle$, $o=m+n$. For a multi-mode moment, the least sampling number for a two-mode moment $\langle(\hat{a}^\dagger)^m\hat{a}^n(\hat{b}^\dagger)^p\hat{b}^q\rangle$ is
\begin{equation}
    N=(1+N_0^{a})^{m+n}(1+N_0^{b})^{p+q}.
    \label{eq:data_noise_relation_multi}
\end{equation}

Here, $N_0^a$ and $N_0^b$ are the effective noise photon numbers in the detection chain for modes $\hat{a}$ and $\hat{b}$, respectively. In our experiment, the effective noise photon number of the $\ket{f0}$--$\ket{g1}$~($\ket{h0}$--$\ket{e1}$) mode is 2.4~(3.5). This is due to the different gains of the JPA for the two modes. 

\subsection{Phase correction of measured signals}
In the experiment, due to the possible phase difference between the qubit drives and $\ket{f0}$--$\ket{g1}$ ($\ket{h0}$--$\ket{e1}$) drives, additional phases might be introduced into the generated states. Here, we used a similar method to Ref.~\citenum{sunada2024efficient}, by maximizing the real part of corresponding moments, the additional phase difference between two modes induced by qubit drives can be canceled. For simplicity, we show the case of the two-logical-qubit state.

For a two-logical-qubit state, it should have four photonic modes, naming their annihilation operators as $\{\hat{a},\hat{b},\hat{c},\hat{d}\}$, respectively.
Thus, from Eq.~\eqref{app_eq:logical_cluster}, the possible additional phase between $\ket{\omega_1}$ and $\ket{\omega_2}$ in the first logical qubit and the second logical qubit can be obtained as:
\begin{equation}
    \begin{split}
        \phi_1&=\mathrm{Arg}\left(\braket{\hat{a}^\dagger\hat{b}\hat{d}^\dagger\hat{d}}-\braket{\hat{a}^\dagger\hat{b}\hat{c}^\dagger\hat{c}\hat{d}^\dagger\hat{d}}\right),\\
        \phi_2&=\mathrm{Arg}\left(\braket{\hat{b}^\dagger\hat{b}\hat{c}^\dagger\hat{d}}-\braket{\hat{a}^\dagger\hat{a}\hat{b}^\dagger\hat{b}\hat{c}^\dagger\hat{d}}\right).
    \end{split}
\end{equation}

After canceling these two phases from all the moments, we start the state reconstruction process based on the method mentioned in Appendix~\ref{subsec:method_tomo} for a four-photon-mode state.

\subsection{State tomography based on local correlations}

For a photon state with large mode numbers, using the method we used in Sec.~\ref{subsec:method_tomo} becomes difficult. From Eq.~\eqref{eq:data_noise_relation_multi}, we find that the amount of sampling required for state reconstruction grows exponentially with the number of modes. For our experiment, a three-logical-qubit state is already reaching the limit, requiring $1\times 10^7$ data samples. Collecting enough data samples for states with more than three logical photon modes is not feasible within a reasonable time. Thus, we used another method for the tomography of photonic states with more than three logical photon modes. This method is based on the fact that the generated state is a type of matrix product state (MPS)~\cite{schon2005sequential}. Thus, measuring local correlations along the state chain allows us to reconstruct the whole state~\cite{sunada2024efficient,baumgratz2013scalable}.

\subsubsection{MPO representation of the generated state}
\begin{figure}
    \centering
    \includegraphics{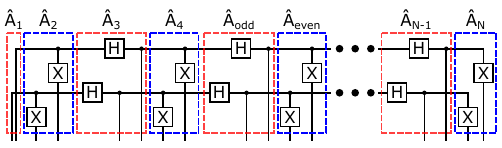}
    \caption{MPO representation of the dual-rail frequency-bin cluster state.}
    \label{app-fig:tensor_rep}
\end{figure}

As mentioned in Sec.~\ref{sec:method}, for generating a logical cluster state, the photon emission process from a transmon qubit is used. If the transmon qubit is prepared in a superposition state, after the photon generation process, the transmon--photon--photon state becomes
\begin{equation}
\alpha\ket{g}+\beta\ket{e} \rightarrow \alpha\ket{g}\otimes\ket{0}\otimes\ket{1}+\beta\ket{e}\otimes\ket{1}\otimes\ket{0}.
\end{equation}
Here, the first photon mode represents the photon mode at the $\ket{h0}$--$\ket{e1}$ transition frequency, and the second one is the photon mode at the $\ket{f0}$--$\ket{g1}$ transition frequency. This photon emission process can be realized by the circuit shown in Fig.~\ref{fig:scheme}(c). Thus, with the transformation used in Ref.~\citenum{biamonte2019lectures}, the MPO representation of the generated state is drawn as Fig.~\ref{app-fig:tensor_rep}.

\subsubsection{MPO reconstruction}
Based on the MPO representation of the state, we can express the state as~\cite{baumgratz2013scalable}
\begin{equation}
    \hat{\rho}=\frac{1}{2^N}\hat{\mathrm{A}}_1\cdots\hat{\mathrm{A}}_N.
\end{equation}
And the corresponding operators are
\begin{align}
        \hat{\mathrm{A}}_1&=\begin{bmatrix}\hat{I}_1,\hat{X}_1,-\hat{Y}_1,\hat{Z}_1\end{bmatrix},\\
        \hat{\mathrm{A}}_{i=\mathrm{even}}&=
        \begin{bmatrix}
        \hat{I} & 0& 0 & -\hat{Z}\\
        0 & \hat{X} &\hat{Y} & 0\\
        0 & -\hat{Y} &\hat{X} & 0\\
        -\hat{Z} & 0& 0 & \hat{I}\\
        \end{bmatrix},\\
        \hat{\mathrm{A}}_{i=\mathrm{odd}}&=\begin{bmatrix}
        \hat{I} & 0& 0 & \hat{Z}\\
        \hat{Z} & 0& 0 & \hat{I}\\
        0 & -\hat{Y} &-\hat{X} & 0\\
        0 & \hat{X} &-\hat{Y} & 0\\        
        \end{bmatrix},\\
        \hat{\mathrm{A}}_N&= \begin{bmatrix}
        \hat{I}_N \\
        \hat{X}_N \\
        -\hat{Y}_N \\
        -\hat{Z}_N \\
        \end{bmatrix}.
    \end{align}
For clarity, we further present the explicit contracted form of the resulting state for four physical modes (corresponding to two logical modes) as an example:
\begin{equation}
\begin{split}
\rho=&\frac{1}{2^4}\hat{\mathrm{A}}_1\hat{\mathrm{A}}_2\hat{\mathrm{A}}_3\hat{\mathrm{A}}_4\\
=&\frac{1}{16}\begin{bmatrix}\hat{I},\hat{X},-\hat{Y},\hat{Z}\end{bmatrix}
\begin{bmatrix}
        \hat{I} & 0& 0 & -\hat{Z}\\
        0 & \hat{X} &\hat{Y} & 0\\
        0 & -\hat{Y} &\hat{X} & 0\\
        -\hat{Z} & 0& 0 & \hat{I}\\
\end{bmatrix}\\
&\begin{bmatrix}
        \hat{I} & 0& 0 & \hat{Z}\\
        \hat{Z} & 0& 0 & \hat{I}\\
        0 & -\hat{Y} &-\hat{X} & 0\\
        0 & \hat{X} &-\hat{Y} & 0\\        
\end{bmatrix}
\begin{bmatrix}
        \hat{I} \\
        \hat{X} \\
        -\hat{Y} \\
        -\hat{Z} \\
\end{bmatrix}\\
=&\frac{1}{16}\bigl(\hat{I}\hat{I}\hat{I}\hat{I}-\hat{Z}\hat{Z}\hat{I}\hat{I}-\hat{I}\hat{I}\hat{Z}\hat{Z}+\hat{Z}\hat{Z}\hat{Z}\hat{Z}\\
&+\hat{X}\hat{X}\hat{Z}\hat{I}-\hat{X}\hat{X}\hat{I}\hat{Z}+\hat{Y}\hat{Y}\hat{Z}\hat{I}-\hat{Y}\hat{Y}\hat{I}\hat{Z}\\
&-\hat{X}\hat{Y}\hat{Y}\hat{X}+\hat{Y}\hat{X}\hat{Y}\hat{X}+\hat{X}\hat{Y}\hat{X}\hat{Y}-\hat{Y}\hat{X}\hat{X}\hat{Y}\\
&-\hat{I}\hat{Z}\hat{X}\hat{X}-\hat{I}\hat{Z}\hat{Y}\hat{Y}+\hat{Z}\hat{I}\hat{X}\hat{X}+\hat{Z}\hat{I}\hat{Y}\hat{Y}
\bigr)
\end{split}
\end{equation}
Using the same method as in Ref.~\citenum{sunada2024efficient}, we find that a five-mode correlation is necessary to reconstruct the whole state. Through the same method as in Eq.~\eqref{eq:bi_mode_moment}, the five-mode correlations can be obtained from the measured voltages. 
Then, with the following target function, the density matrix of the states with more modes can be reconstructed without increasing the sampling number. 
\begin{widetext}
\begin{equation*}
\begin{split}
    L=-\sum_{m,n,p,q,i,j,k,l,u,v}^{s=1,\ldots,N-4}\frac{1}{\delta_{m,n,p,q,i,j,k,l,u,v}^2}&
    \times\biggl|\langle(\hat{a}_{s}^\dagger)^m\hat{a}_{s}^n
    (\hat{a}_{s+1}^\dagger)^p\hat{a}_{s+1}^q
    (\hat{a}_{s+2}^\dagger)^i\hat{a}_{s+2}^j
    (\hat{a}_{s+3}^\dagger)^k\hat{a}_{s+3}^l
    (\hat{a}_{s+4}^\dagger)^u\hat{a}_{s+4}^v\rangle\\
    &-\mathrm{Tr}\left(\hat{\rho}(\hat{a}_{s}^\dagger)^m\hat{a}_{s}^n
    (\hat{a}_{s+1}^\dagger)^p\hat{a}_{s+1}^q
    (\hat{a}_{s+2}^\dagger)^i\hat{a}_{s+2}^j
    (\hat{a}_{s+3}^\dagger)^k\hat{a}_{s+3}^l
    (\hat{a}_{s+4}^\dagger)^u\hat{a}_{s+4}^v\right)\biggr|^2.
\end{split}    
\end{equation*}
\end{widetext}

Here $\hat{a}_s^\dagger$ and $\hat{a}_s$ are the creation and annihilation operators of mode $s$. Instead of using the correlations based on Pauli operators as in Ref.~\citenum{sunada2024efficient}, we use the moments of the measured modes. Although they differ in form, the information they contain is consistent. Under the single-photon assumption, they can be converted into each other. Therefore, the discussion in Ref.~\citenum{sunada2024efficient} remains applicable here.

\subsection{Process tomography\label{subsec:method_process_tomo}}
\begin{figure}
    \centering
\includegraphics{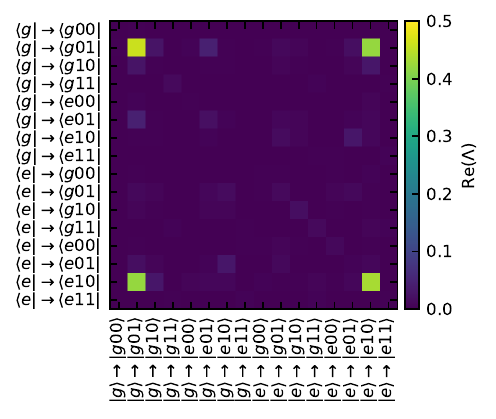}
    \caption{Real part of the Choi matrix $\Lambda$ of the two-photon emission process.}
    \label{app_fig:choi_matrix}
\end{figure}

We have also applied full-state tomography on the qubit-emitted photon-pair system to obtain the Pauli transfer matrix of the whole photon emission process. We prepare the qubit state in the following six initial states $\{\ket{g},\ket{e},\ket{g}\pm\ket{e},\ket{g}\pm i \ket{e}\}$. We then apply the photon emission process for each initial state and measure the corresponding complex amplitude of emitted photon pairs together with the qubit state. For the qubit measurement, we apply different qubit rotations before the Pauli matrices measurement of the qubit in the basis $\{\hat{\sigma}_I,\hat{\sigma}_x,\hat{\sigma}_y,\hat{\sigma}_z\}$. With the following target function,
\begin{equation}
\begin{split}
    L=&-\sum_{m,n,p,q,i}\frac{1}{\delta_{m,n,p,q,i}^2}\\
    &\times\left|\langle(\hat{a}^\dagger)^m\hat{a}^n(\hat{b}^\dagger)^p\hat{b}^q\hat{\sigma}_i\rangle-\mathrm{Tr}\left(\hat{\rho}(\hat{a}^\dagger)^m\hat{a}^n(\hat{b}^\dagger)^p\hat{b}^q\hat{\sigma}_i\right)\right|^2,
\end{split}    
\end{equation}
the qubit-emitted photon pairs system for different qubit initial states can be reconstructed. Then, with these results, the Pauli transfer matrix of the emission process is reconstructed, and the process fidelity is calculated. We can also predict states with more modes by repeating the emission process. Figure~\ref{app_fig:choi_matrix} shows the Choi matrix $\Lambda$ of the photon emission process~\cite{schwartz2016deterministic,besse2020realizing},
\begin{equation}
\begin{split}
\left(\alpha\ket{g}+\beta\ket{e}\right)\ket{0}&\rightarrow\alpha\ket{g}\ket{\omega_1}+\beta\ket{e}\ket{\omega_2}\\
&\coloneqq\alpha\ket{g}\ket{01}+\beta\ket{e}\ket{10}.
\end{split}
\label{app_eq:choi_two_freq_generate}
\end{equation}
The dashed lines in Figs.~\ref{fig:fidelity_PTM} and~\ref{fig:fig_LE} are calculated based on this Choi matrix.

 \section{Localizable entanglement\label{app_sec:LE}}
Localizable entanglement~(LE) has been used to estimate how much entanglement remains between different modes of the generated state~\cite{verstraete2004entanglement,popp2005localizable,schwartz2016deterministic}. The LE between two modes can be calculated by the following steps.
\begin{enumerate}
\item First, we apply local projective measurements to all other modes. 
\item Then, we calculate the negativity within the reduced two-mode system.
\item LE is defined as the expectation value of negativity over all the possible projective measurement outcomes.
\end{enumerate}
 
We need to choose projection operators to calculate the LE between modes. In this section, we will explain how to choose the projection operators for the dual-rail cluster states we generated based on the case of an ideal state. This choice gives us a good lower bound of the LE for the non-ideal state we generated in the experiment~\cite{sunada2024efficient}.

\begin{figure}
    \centering
\includegraphics{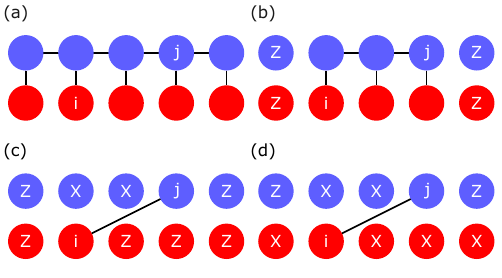}
    \caption{Protocol to calculate the localizable entanglement between two modes $i$ and $j$ in a five-logical-qubit dual-rail cluster state. (a)~Two modes chosen. (b)~Graph of the state after removing the modes outside the target modes in the graph representation by applying Pauli-Z operators to them, as mentioned in Step 1. (c)~Graph of the state after applying Pauli operators mentioned in Step 2. Two target modes are linked to each other. (d)~Actual Pauli operators we applied in the data processing, as mentioned in Step 3.}
    \label{app_fig:LE_cal}
\end{figure}

In Sec.~\ref{sec:method}, we showed that the states we generated are equivalent to the graph representation with local operations as in Fig.~\ref{fig:scheme}(e). Therefore, to calculate the LE, we start from the graph representation of this state.

In Pauli measurements on the graph states, each Pauli operator has a different effect on the state graph. A Pauli-$Z$ operator deletes the corresponding vertex~(mode) $i$ from the whole graph~(state). Pauli-$X$ and Pauli-$Y$ operators, in our cases, locally complement the neighborhood of the corresponding vertex~(mode) $i$ and delete it from the whole graph~(state)~\cite{hein2004multiparty}.

Therefore, when we calculate the LE directly between two modes, for example, two modes $i$ and $j$ in a five-logical-qubit state as shown in Fig.~\ref{app_fig:LE_cal}(a), we follow the steps below.

\begin{enumerate}
    \item As a first step, we note that the frequency-bin dual-rail cluster state we generate is locally equivalent to the graph state shown in Fig.~\ref{app_fig:LE_cal}(a), by applying Hadamard gates to all qubits in the second row. The following procedure is performed on this equivalent graph state.
    \item First, we apply Pauli-$Z$ operators to all vertices outside the vertical lines where the two vertices we want to measure are located, as shown in Fig.~\ref{app_fig:LE_cal}(b).
    \item Then, for the remaining vertices, we apply Pauli-$X$ operators to the vertices in the first row and Pauli-$Z$ operators to the vertices in the second row, as shown in Fig.~\ref{app_fig:LE_cal}(c).
\item Finally, to translate the results back to the original frequency-bin dual-rail cluster state, we conjugate the Pauli-$Z$ operators applied to the second-row qubits into Pauli-$X$ operators by Hadamard gates, as shown in Fig.~\ref{app_fig:LE_cal}(d).

\end{enumerate}

The localizable entanglements of all the states in the physical space in Figs.~\ref{fig:fig_LE} and~\ref{app_fig:LE}, are calculated in this way. For the localizable entanglements in the logical subspace, we chose the operators used for a linear cluster state~\cite{sunada2024efficient}.
 \section{Additional plots\label{app_sec:App_exp}}
\subsection{Photon state tomography}

\begin{figure*}
    \centering
\includegraphics{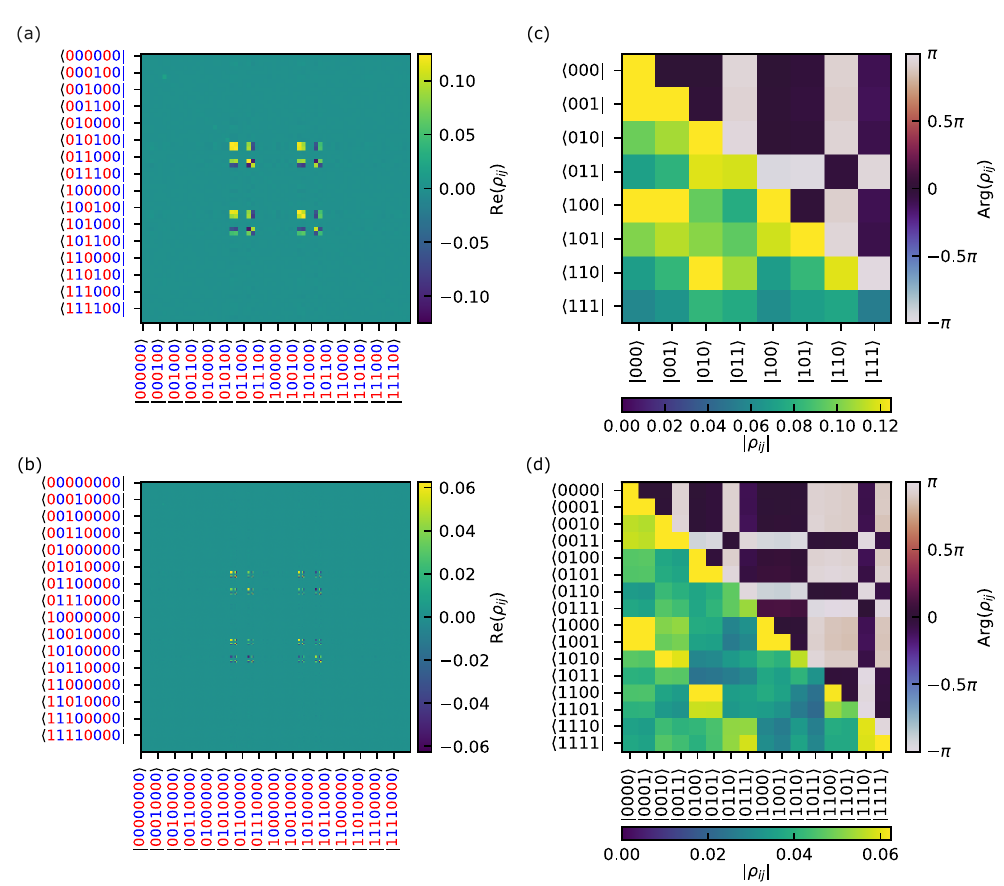}
\caption{Reconstructed density matrices of the generated three- and four-logical-qubit frequency-bin cluster states. (a),(b)~Real part of the density matrices of the reconstructed states. The blue (red) colored modes in the bras and kets correspond to the frequency bin of $\ket{\omega_1}(\ket{\omega_2})$. Each adjacent pair (i.e., at positions 0 and 1, 2 and 3, etc.) corresponds to two modes within the same time bin. (c),(d)~Density matrices of the reconstructed states, projected onto the logical subspace spanned by $\{\ket{\omega_1},\ket{\omega_2}\}$. Absolute values are plotted in the diagonal and the lower-left triangle, and the complex arguments are plotted in the upper-right triangle.
}
    \label{app_fig:app_tomo}
\end{figure*}

Here we show the reconstructed density matrix of the generated three- and four-logical-qubit states in Fig.~\ref{app_fig:app_tomo}.

\subsection{Localizable entanglement}
\begin{figure}
    \centering
\includegraphics{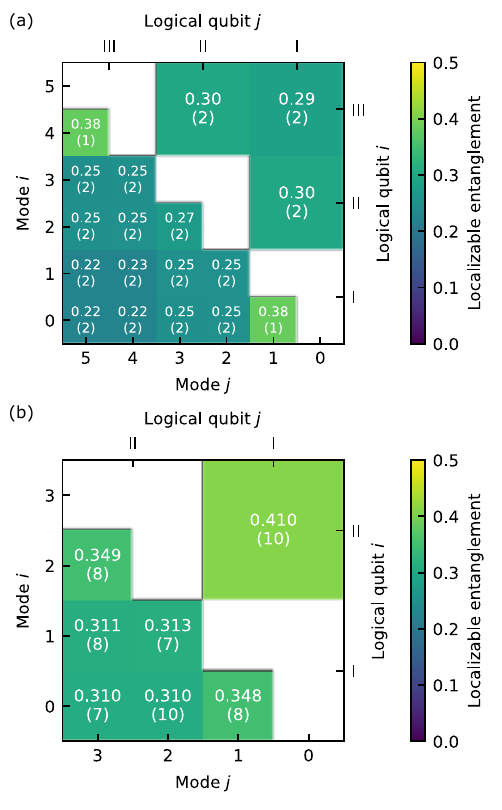}
    \caption{Localizable entanglement~(LE) and its standard deviation between different modes in the generated states. (a)~LE in a three-logical-qubit state. (b)~LE in a two-logical-qubit state. The number in parentheses indicates the one-standard-deviation statistical uncertainty, referring to the uncertainty in the last digit(s) of the LE value. The lower-left part of the figure shows the LE between the physical modes, while the upper-right part corresponds to the LE between the corresponding logical qubits.}
    \label{app_fig:LE}
\end{figure}

Here we show the results of LE for the generated three-, and four-logical-qubit states in Fig.~\ref{app_fig:LE}.

\section{Numerical simulation\label{app_sec:Theo_simu}}
\subsection{Error budget}
For our protocol, as mentioned in the main text, the two main limitations are the overlap between the two frequency modes due to limited detuning, and the finite coherence time of the qubit. Although these effects are difficult to separate experimentally, we perform two independent simulations to evaluate their respective impacts.

\subsubsection{Coherence limit\label{app_subsubsec:coherence_error}}
To examine how the qubit energy relaxation and dephasing affect the state-generation process, and how our frequency-encoding scheme mitigates these effects, we consider a simplified Hamiltonian describing ideal photon emission between the qubit and two orthogonal modes:

\begin{equation}
\hat{H}_{s}/\hbar=g_{\mathrm{eff}}^{(1)}\ket{f00}\bra{g01}+g_{\mathrm{eff}}^{(2)}\ket{h00}\bra{e10}+h.c.
\end{equation}

\begin{figure}
    \centering
    \includegraphics{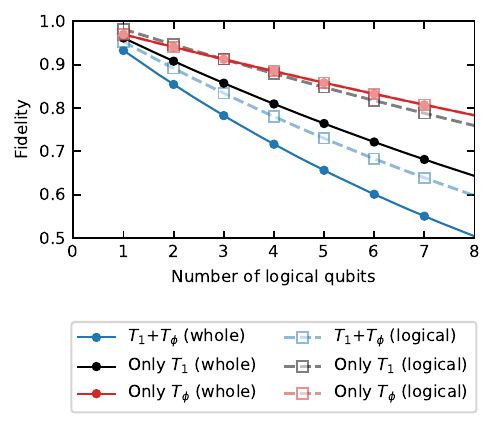}
    \caption{Fidelity of the generated cluster state as a function of the number of logical qubits, evaluated under different error models: (i) including both qubit energy relaxation ($T_1$) and pure dephasing ($T_\phi$) processes (blue), (ii) including only energy relaxation ($T_1$) (black), and (iii) including only dephasing ($T_\phi$) (red). For each case, the fidelity is evaluated over the entire physical Hilbert space (“whole”, solid circles and solid lines) and within the logical subspace (“logical”, open squares and dashed lines). Because pure dephasing acts identically on both subspaces, the two corresponding traces are fully overlapping.}
    \label{fig:fig_coherence_error}
\end{figure}

Here, both coupling strengths are set to $g_{\mathrm{eff}}^{(1)}/2\pi=g_{\mathrm{eff}}^{(2)}/2\pi=0.5$~MHz, such that after $1~\mathrm{\,\mu s}$ (the pulse length we used in the experiment), the qubit population is completely transferred to the two orthogonal frequency modes. In this idealized case, any effect from the spectral overlap between the modes is eliminated. We simulate the photon-generation process under the following conditions:
\begin{enumerate}
    \item Both energy decay rates $\Gamma_D^{i}$ and dephasing rates $\Gamma_\phi^i$ are included;
    \item Only energy decay rates $\Gamma_D^{i}$ are included (no pure dephasing);
    \item Only dephasing rates $\Gamma_D^{i}$ are included (no energy decay).
\end{enumerate}
Here $\Gamma_D^i=1/T_i^i$ and $\Gamma_\phi^i=(1/T_2^i-1/2T_i^i)/2$ ($i=ge,ef,fh$). We assume that the energy decay time and decoherence time are normally distributed, with the means and standard deviations obtained from the experimental results (see Table~\ref{tab:sys_parameter}). The simulation results are summarized in Fig.~\ref{fig:fig_coherence_error} and Table~\ref{tab:simu_results}.

 \begin{table}
    \caption{Photon-generation performance under different decoherence conditions.}
    \centering
    \begin{ruledtabular}
    \begin{tabular}{lccc}
    &Cond.~1&Cond.~2&Cond.~3\\
    \hline
    $F_p$ & $91.5\pm1.0\%$ & $94.4\pm1.0\%$ &$97.0\pm0.6\%$\\
    $L_F^{\mathrm{w}}$& $11\pm1$ & $17\pm3$ & $33\pm7$\\
    $L_F^{\mathrm{l}}$& $15\pm2$ & $29\pm6$ & $33\pm7$\\
\end{tabular}
    \end{ruledtabular}
    \label{tab:simu_results}
\end{table}

As shown in Fig.~\ref{fig:fig_coherence_error}, the fidelities of the generated states are compared for the whole space and within the logical subspace. Energy relaxation (Cond.~2) is found to be the dominant source of infidelity, whereas pure dephasing (Cond.~3) mainly affects the coherence without being correctable by our encoding scheme. The improvement observed under Cond.~2 demonstrates that the dual-rail frequency encoding effectively mitigates photon-loss–type errors, whereas pure dephasing errors remain uncorrected.

To quantify the overall performance, we define the fidelity length $L_F^{\mathrm{w(l)}}$ by fitting the fidelity decay with the number of generated logical modes $N$ using an exponential model $F=\exp(-N/L_F^{\mathrm{w(l)}})$. The extracted values confirm that energy relaxation is the primary error source, and that the dual-rail scheme provides significant protection against it but not against dephasing. This behavior arises because energy decay events during photon generation manifest as photon-loss errors, which the encoding can correct, whereas dephasing merely suppresses the off-diagonal coherence terms without producing a detectable photon-loss signature. Moreover, even under Cond.~2, the fidelity of the corrected state cannot reach $100\%$, since a finite energy lifetime $T_1$ necessarily imposes a finite coherence lifetime $2T_1$, leaving residual errors within the logical subspace.

We also simulated performance when the system is under the Purcell limit, and get a process fidelity of $97.2\%$ and fidelity length $L_F^{\mathrm{w}}=36$ and $L_F^{\mathrm{l}}=58$.

\subsubsection{Photon modes overlap\label{app_subsubsec:overlap_error}}
In addition to qubit coherence, another major source of error arises from the finite detuning between frequency modes. In our protocol, two photon modes $\ket{\omega_1}$ and $\ket{\omega_2}$ are not perfectly orthogonal due to the limited frequency detuning. This non-orthogonality leads to a finite temporal overlap between modes, which inherently limits the achievable process fidelity even in the absence of qubit decoherence.

To quantify this effect, we perform a numerical simulation of the photon generation process following the approach in Ref.~\citenum{kiilerich2020quantum}. Considering the higher-order nonlinearity in the transmon qubit, we use a Hamiltonian with a 6th-order nonlinear term, as Eq.~\eqref{app_eq:Hamil} shows. The $\ket{f0}$--$\ket{g1}$ and $\ket{h0}$--$\ket{e1}$ drives are added as the following additional drive terms:
\begin{equation}
\begin{split}
\hat{H}_{\mathrm{d}}/\hbar&=\Omega_{f0g1}\cos(\omega_\mathrm{d}^{f0g1}t)(\hat{b}+\hat{b}^\dagger)\\
&+\Omega_{h0e1}\cos(\omega_\mathrm{d}^{h0e1}t)(\hat{b}+\hat{b}^\dagger).
\end{split}
\end{equation}

\begin{figure}
    \centering
    \includegraphics{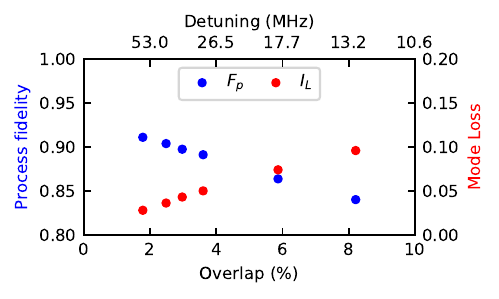}
    \caption{Photon-generation performance under different mode overlap: process fidelity (blue, left axis) and overlap between modes (red, left axis)}
    \label{fig:fig_overlap_error}
\end{figure}

The mode functions of the output modes $\ket{\omega_1}$ and $\ket{\omega_2}$ can be directly extracted from the temporal modes of the output photon $(\ket{0}+\ket{1})/\sqrt{2}$ at each frequency, instead of obtaining them from the correlation function of the output modes like in Ref.~\citenum{kiilerich2020quantum}, which would effectively apply an orthogonalization and lose all the effect of the mode overlap. For the qubit loss operators, here we assume that there is no other energy decay or dephasing, resulting in the qubit being under the Purcell limit of the system. Under these assumptions, we obtain the fidelity of this two-photon generation process under different frequency detunings (mode overlaps) between $\ket{\omega_1}$ and $\ket{\omega_2}$, as shown in Table~\ref{tab:simu_results_overlap} and Fig.~\ref{fig:fig_overlap_error}. In parallel, we can also calculate the photon-mode loss $I_L$ caused by choosing non-orthogonal modes $v_1(t)$ and $v_2(t)$. 

 \begin{table}
    \caption{Photon-generation performance under different mode overlap.}
    \centering
    \begin{ruledtabular}
    \begin{tabular}{ccccccc}
    $|\omega_1-\omega_2|$~(MHz) &12.5 &17.5 & 27.5& 32.5 & 37.5 &47.5\\
Overlap ($\%$) & $8.21$ &$5.86$ & $3.60$ &$2.97$& $2.48$ &$1.78$ \\
    \hline
    $F_p$ & $84.0\%$ & $86.4\%$ & $89.2\%$ &$89.7\%$& $90.4\%$ &$91.1\%$\\
$I_L$ & $9.6\%$ & $7.4\%$ & $5.0\%$ &$4.3\%$& $3.6\%$ &$2.8\%$\\
    $L_F^{\mathrm{w}}$& $5.6$ & $6.6$ & $8.1$ & $8.2$  &  $9.0$ & $9.6$  \\
    $L_F^{\mathrm{l}}$& $15.3$ & $15.6$ & $15.9$ &  $14.8$ & $15.7$ & $15.5$\\
    \end{tabular}
    \end{ruledtabular}
    \label{tab:simu_results_overlap}
\end{table}

As shown in Fig.~\ref{fig:fig_overlap_error}, within the examined range of moderate overlap, the photon-generation fidelity decreases with increasing overlap, whereas the photon-mode loss increases accordingly. Interestingly, we find that the sum of $F_p$ and $I_L$ is close to the Purcell-limited fidelity ($97.2\%$), indicating that $I_L$ can be regarded as one of the main error sources induced by the non-orthogonal modes $v_1(t)$ and $v_2(t)$. We also fitted fidelity length $L_F^{\mathrm{w(l)}}$ for the generated states, as shown in Table~\ref{tab:simu_results_overlap}. From these results, we find that our dual-rail encoding can partially mitigate the errors induced by mode overlap. This indicates that such an overlap causes part of the generated state to leak out of the encoded logical subspace (e.g., into the vacuum or doubly excited states), which can then be corrected. However, the remaining portion of the overlap-induced error manifests as a reduction in coherence within the logical subspace itself and therefore cannot be fully corrected.
This explains the residual difference between the Purcell limit and the sum of $F_p+I_L$

\subsubsection{Total simulation}
After all, we simulate the system based on the method used in Sec.~\ref {app_subsubsec:overlap_error} with loss operators of the qubit. We generate 10 random samples and perform simulations to account for the uncertainty in these parameters. Under these assumptions, we obtain the fidelity of this two-photon generation process as $86.3\pm1.7\%$, with $L_F^{\mathrm{w}}=6.4\pm0.2$ and $L_F^{\mathrm{w}}=11\pm1$. Based on the results shown in Table~\ref{tab:simu_results} and Table~\ref{tab:simu_results_overlap}, we estimated that the qubit energy-decay induced an infidelity of around $4.5\sim6.5\%$, the qubit decoherence induced an infidelity of around $2.4\sim3.6\%$, and the mode overlap induced an infidelity of around $6\sim7\%$ during each photon generation process. And by defining an effective photon generation fidelity per round as $F_{\mathrm{eff}}^l=\exp(-1/L_F^l)$,  we estimate that after the error correction, the qubit energy-decay induced infidelity can be decreased to around $3.5\sim4\%$ and the mode overlap induced infidelity can be decreased to around $3\sim4\%$ per round, while the infidelity induced by the qubit decoherence remains.

\subsection{Performance comparison}
\begin{figure}
    \centering
\includegraphics{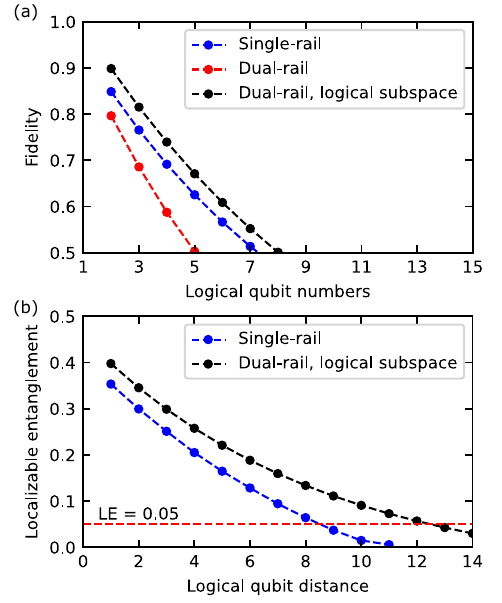}
    \caption{Comparison of the cluster-state fidelity and localizable entanglement~(LE) between frequency-bin dual-rail encoding and conventional single-photon (single-rail) encoding schemes. The black and red dots represent the simulated results for the frequency-bin encoding used in this experiment, while the blue dots correspond to simulated results for single-photon encoding using the same device parameters. (a)~Fidelity of the generated cluster state in the logical subspace as a function of the number of logical qubits. Here, we also show the fidelity of the dual-rail state in the whole space as the red dots. (b)~LE extracted from the same states.}
    \label{fig:fig_comparation}
\end{figure}

As we mentioned in Sec.~\ref{sub_sec:exp_performance_comp}, we also conducted a numerical simulation of a conventional single-photon (single-rail) encoding scheme using the same device parameters. For this simulation of single-rail encoding, the mode function is also obtained from the temporal mode of the output photon $(\ket{0}+\ket{1})/\sqrt{2}$ in mode $\ket{\omega_1}$, while the other conditions are not changed. The results are shown in Fig.~\ref{fig:fig_comparation}. Our findings show that, with frequency-bin encoding, the fidelity of the generated cluster state in the logical subspace remains above $50\%$ for up to 8 logical qubits, whereas the single-photon encoding maintains this threshold only up to 7 modes, as shown in Fig.~\ref{fig:fig_comparation}(a). Similarly, the LE length in our frequency-encoded states reaches 12, compared to 8 in the single-photon case, as shown in Fig.~\ref{fig:fig_comparation}(b). The frequency-bin encoding shows a significantly longer LE length, indicating more robust entanglement under the same experimental constraints. 

\FloatBarrier
\bibliography{bibliography}  

\end{document}